\documentstyle[12pt]{article}

\def\beq{\begin{equation}}
\def\eeq{\end{equation}}
\def\eq{\end{equation}}

\def\({\left (}
\def\){\right )}

\def \M{{\cal M}}
\def\mgravitino{m_{3/2}}

\def\be{\bar{e}}

\def\bu{\bar{u}}

\def\bd{\bar{d}}

\def\bj{\bar{j}}

\def\bphi{\bar{\phi}}

\def\nphi{n_{\phi}}
\def\mphi{m_{\phi}}
\def\rhophi{\rho_{\phi}}

\def\K1{K^{(1)}}
\def\intD{\int d^4 \theta}

\def\({\left (}
\def\){\right )}

\def\begi{\begin{itemize}}
\def\endi{\end{itemize}}

\begin{document}

\begin{titlepage}

\begin{flushright}
SLAC-PUB-95-6846 \\
MIT-CTP-2441\\
SCIPP 95/20\\
hep-ph/9507453
\end{flushright}



\LARGE
\vspace{0.3in}

\begin{center}
{\bf Baryogenesis from Flat Directions of the Supersymmetric
Standard Model  }
\vspace{0.4in}

\normalsize

Michael Dine\footnote{Work supported in part by the Department
of Energy.}\\
\vspace{.05in}
{\it Santa Cruz Institute for Particle Physics \\
University of California \\
Santa Cruz, CA 95064}

\vspace{.15in}

Lisa Randall\footnote{Work
supported in part by the Department of Energy under contract
DE-AC02-76ER03069, and NSF grant PHY89-04035.
NSF Young Investigator Award, Alfred Sloan
Foundation Fellowship, and DOE Outstanding Junior Investigator Award.
}\\
\vspace{.05in}
{\it Laboratory for Nuclear Science and Department of Physics \\
Massachusettts Institute of Technology \\
Cambridge, MA 02139}

\vspace{.15in}

Scott Thomas\footnote{Work supported by the Department of Energy
under contract DE-AC03-76SF00515.}\\
\vspace{.05in}
{\it Stanford Linear Accelerator Center\\
Stanford University\\
Stanford, CA 94309}

\end{center}

\vspace{0.2in}

\vspace{0.25in}

\newpage


\LARGE
\begin{center}
Abstract
\end{center}
\normalsize

Baryogenesis from the coherent production of a scalar condensate along
a flat direction  of the supersymmetric extension of the standard model
(Affleck-Dine mechanism) is investigated.  Two important effects are
emphasized.  First, nonrenormalizable terms in the superpotential can
lift standard model flat directions at large field values.  Second,
the finite energy density in the early universe induces soft potentials
with curvature of order the Hubble constant.  Both these have important
implications for baryogenesis, which requires large squark or slepton
expectation values to develop along flat directions.  In particular,
the induced mass squared must be negative.  The resulting baryon to
entropy ratio is very insensitive to the details of the couplings and
initial conditions, but depends on the dimension of the nonrenormalizable
operator in the superpotential which stabilizes the flat direction
and the reheat temperature after inflation.  Unlike the original
scenario, an acceptable baryon asymmetry can result without subsequent
entropy releases.  In the simplest scenario the baryon asymmetry is
generated along the $LH_u$ flat direction, and is related to the mass
of the lightest neutrino.

\end{titlepage}

\section{Introduction}

One of the features which distinguishes supersymmetric
field theories from ordinary ones is the existence of
``flat directions" in field space on which the scalar potential vanishes.
At the level of renormalizable terms, such flat directions
are generic.
Supersymmetry breaking lifts these directions and
sets the scale for their potential.
{}From a cosmological perspective, these flat directions
can have profound consequences.  The parameters which
describe the flat directions can be thought of as expectation
values of massless chiral fields (moduli).
These expectation values can start out displaced
from the true minimum.
Oscillations about the minimum occur when
the Hubble constant
becomes comparable to the effective mass.
These oscillations
have an equation of state without pressure, and so
amount to a coherent condensate of zero-momentum particles,
redshifting like matter.
The coherent production of scalar fields along flat directions
emerges as a generic feature of  any supersymmetric theory.

String theories contain a number of perturbative flat directions
whose potential is generated in the presence of supersymmetry breaking.
At early times the moduli can in general
have Planck scale expectation values relative to the true minimum.
Because these fields probably decay only through Planck suppressed
interactions, they would
dominate the energy density of the universe before
decaying. This potential
cosmological catastrophe is the string
version \cite{bkn} of the ``Polonyi problem" \cite{polonyi}.

The main focus of this paper will be on another class
of flat directions, which are present
in the minimal supersymmetric standard model (MSSM).
In these directions combinations
of fields carrying baryon or lepton
number, such as squarks
and sleptons, have non-zero expectation values.
If baryon and lepton number are explicitly broken
it is possible to excite a non-zero baryon or lepton number along such
directions, as first suggested by Affleck and Dine \cite{ad}.
We will refer to this as the Affleck-Dine (AD)
mechanism of baryogenesis, and the
associated fields as AD fields.  Eventually,
this condensate decays leaving the universe
with a non-zero baryon or lepton number.

In this paper, we carefully re-examine the coherent production
of scalar fields along flat directions.
In order to make any quantitative predictions
about the production of coherent condensates in the early
universe, a number of important questions must be faced.
First,
how are the flat directions lifted in the early universe?
Second,
for the AD mechanism, how
do the baryon number violating terms in the potential arise?
And third,
what are the ``initial conditions'' for the flat directions?
These are the questions which we wish to address.

We begin by observing that
the finite energy density of the early universe,
breaks supersymmetry by a ``large'' amount \cite{adshort,
hubblemass}.
Gravitational strength interactions
between the background energy density and flat directions
give soft supersymmetry
breaking terms, with scales which are parametrically
of order the Hubble constant.
At very early times it is this breaking which is most important
in lifting
the flat directions.
This is in contrast to the usual assumption that
the potential for flat directions   arises only from
the   supersymmetry breaking terms that determine the
soft potential in the present universe.
According to the standard picture (with Hubble
scale masses neglected), fields are effectively frozen
in the early universe (up to quantum deSitter fluctuations
during inflation) and highly overdamped.
However, the supersymmetry breaking due
to the finite energy density
gives rise to masses of order $H$. The scalar
field is therefore parametrically close
to critically damped at early times, and
can efficiently evolve to an
instantaneous minimum of the potential.
This qualitatively changes the scenario for the coherent
production of scalar fields as discussed below.

In the case of the moduli problem,
the existence of a nontrivial
potential at early times with a minimum which
does not necessarily coincide with the minimum
at late times just
brings the cosmological problem into sharper focus, and calls into
question some proposed solutions.  It also suggests
a possible solution if the minimum of the moduli
potential at early and late times
coincides.
This can be technically natural
if the minima lie at a point of enhanced symmetry.  We will
discuss briefly this solution and its principle limitation:
the dilaton.

The finite density supersymmetry
breaking potential also has an important impact on the
AD mechanism.
A flat direction can easily have a minimum
far from the origin, giving rise to large
expectation values for squark and/or slepton fields.
 On the other hand, because
the curvature of the potential is of order the
(instantaneous) Hubble constant, if the minimum
of the potential is at the origin,
quantum fluctuations of this field during inflation
will not lead to a net baryon numbe (as is frequently
assumed) since
the correlation volume for the fluctuations
is generally much smaller than the present Hubble volume.

There is another important issue which must be dealt
with in the case of the AD field:  the flat
directions are expected to be lifted by higher dimension operators.
This has two effects.
First, the expectation value along a flat direction
at early times is determined by a balance between the induced
soft potential and the higher order superpotential terms, and
is typically small compared to the scale
of the higher dimension operators.
On the other hand, these operators themselves
generally violate baryon and lepton number.
As a result,
the baryon number per particle in the
condensate is order one.
However, the fraction of the
energy (and finally entropy) density carried by
the condensate is typically rather small,
and a sensitive function of the dimension of these
operators.
Because of this it is possible for the AD mechanism to produce
the correct baryon to entropy ratio even without additional
entropy releases, as opposed to the usual claims in the
literature.
In the end, we find that the
mechanism is quite robust.  There exist many flat directions
which are broken only by operators of  high dimension,
and for which an acceptable baryon number is obtained.
As for the decay of the AD field, we find that
generally it evaporates by scattering with the thermal plasma,
not by free decay as assumed in the early literature.
It is also possible for the AD direction to be exactly flat in the
supersymmetric limit,
and we will discuss this possibility as well.

The finite density supersymmetry breaking also helps to answer
the question of initial conditions along flat directions,
which are usually assumed in some ad hoc way.
%
In a cosmological scenario which includes inflation,
the conditions relevant here are simply the
values of the scalar fields along the flat directions at the
end of inflation.
During inflation
the finite vacuum energy breaks supersymmetry
and generates a soft potential.
If the duration of inflation is sufficient to solve the flatness
and horizon problems,
the flat directions are efficiently driven toward
an instantaneous minimum of the
potential.
This sets the ``initial conditions'' for the subsequent
evolution.

The structure of the paper is as follows.
In the following section the relevant properties of flat directions
are reviewed.
In section 3, supersymmetry breaking in the early universe is discussed,
and the supergravity interactions responsible for transmitting
this breaking to the flat directions presented.
We show that within a cosmological scenario
with an inflaton with  sufficiently low reheat temperature
to evade the gravitino problem, the finite energy
supersymmetry breaking is important during inflation and the inflaton
matter dominated era following inflation.
In section 4 the impact on the AD mechanism
is discussed.
The classical evolution along a flat direction is studied
for the case of a negative soft mass squared.
The possibility of realizing the AD
mechanism with a small positive mass squared is also considered.
The Polonyi problem is reconsidered in the light of the finite
energy density supersymmetry breaking in section 5.
The appendix contains a list of flat directions for
the standard model.

Throughout we assume for simplicity a hidden sector
model of supersymmetry breaking in which the zero density
breaking is transmitted to the visible sector by gravitational
scale interactions.
The gravitino mass, $\mgravitino$, then sets the weak scale,
and is related to the intermediate scale
of supersymmetry breaking by
$\mgravitino \sim M_{{\rm INT}}^2 / M_p$.

\section{Flat Directions}

Supersymmetric theories commonly have directions
with no classical potential.
The space of all flat directions is usually referred to as the
moduli space.
In string theory moduli fields which parameterize an internal
conformal field theory are common.
In some cases the degeneracy for these moduli arises from
a world sheet symmetry.
In other cases it can be understood in terms of space time
discrete $R$ symmetries \cite{miracles}.
More generally flat directions are common in supersymmetric
field theories, particularly ones with a large number
of fields, such as the MSSM.
In this case flat directions arise as accidental
degeneracies along which both $D$ and $F$ components vanish.
In this section we review some properties of flat directions
which are important for the AD mechanism.
We also discuss the effects which lift the flat directions,
namely higher dimension operators and soft supersymmetry
breaking terms.

The classical degeneracy along flat directions is protected from
perturbative quantum corrections in the
supersymmetric limit by the nonrenormalization theorem \cite {nrt}.
The degeneracy can be lifted by nonperturbative quantum corrections.
For the flat directions relevant to the
AD mechanism these effects are unimportant since no visible sector
gauge couplings become strong in the early universe.
We will therefore assume the potential vanishes on the
moduli space in the supersymmetric  and $M_p \to \infty$ limit.
The potential then appears as a result of supersymmetry
breaking and nonrenormalizable terms in the
superpotential.

A flat direction is parameterized by a full chiral
superfield, including scalar, fermionic, and auxiliary
components.
Here, however, the term ``flat direction" will usually
refer only to the scalar component as
we are interested in the coherent production of scalar
fields.
A single flat direction necessarily carries a global
$U(1)$ quantum number.
A condensate of the flat direction can therefore carry
a net particle number for some $U(1)$,
as required for the AD mechanism.
As discussed in subsequent sections,
condensates of standard model flat directions
turn out to decay in the early universe through renormalizable
couplings when the temperature is well above the weak scale.
At such temperatures anomalous sphaleron processes
which violate $B+L$ are
in equilibrium \cite{krs}.
The relevant quantum number the condensate must carry
in order to give a nonvanishing $B$ after sphaleron
processing is therefore $B-L$.
The minimal standard model contains a large number of directions
which are flat with respect to the renormalizable interactions
and carry $B-L$.
The subspace on which the gauge potential arising from
$D$ terms vanishes is 37 dimensional.
There are a large number
of directions in this subspace for
which all $F$ components also vanish.
A typical example of a renormalizable
flat direction, carrying $B-L=-1$, is
$$
Q^{\alpha}_1 = {1\over \sqrt{3}}\left( \begin{array}{c} \phi \\ 0 \end{array}
\right)~~~~~~
L_1 = {1\over \sqrt{3}}\left( \begin{array}{c} 0 \\ \phi \end{array}
\right)~~~~~~
\bar{d}^{\alpha}_2 = {1\over \sqrt{3}}\phi
$$
where superscripts are color indices,
subscripts are for generation,
and $\phi$ is the complex field parameterizing the flat
direction (with canonical kinetic term).
A list of standard model flat directions is
given in the appendix.

It is often convenient to characterize a flat direction by
a composite gauge
invariant operator, $X$, formed from the product of $m$
chiral superfields which make up the direction.
For example, the direction given above may be parameterized
by the invariant $X=Q_1 L_1 \bar{d}_2$ ($m=3$).
The scalar component of the composite operator
is related to the canonical field, $\phi$,
parameterizing the flat direction
by a relation of the form $X=c\phi^m$.
Fields can take on nonzero values along multiple flat directions
simultaneously,
although $F$ flatness is
then generally not maintained for other directions.
For example, the renormalizable $F$ terms vanish if both
$\bu_1 \bd_1 \bd_2$ and $Q_3 L_1 \bd_1$ are nonzero,
while $F_{H_d}^*$ does not vanish if
$\bu_1 \bd_1 \bd_2$ and $Q_2 L_1 \bd_1$ are nonzero.
For multiple flat directions the relation between the
invariant composite operators and the $\phi$ fields with
canonical normalization is in general highly nonlinear.
However most of the relevant dynamics for the AD
mechanism does not require the treatment of multiple directions.
In fact, even when there are multiple flat directions,
it is the lowest dimension operator generating a potential
for one of the flat directions which determines
the ultimate baryon to entropy ratio.
Unless stated otherwise
we therefore consider the dynamics of a single flat direction
in what follows.

Typically, several supermultiplets gain mass in a flat direction,
$m_{\perp} = \lambda \langle \phi \rangle$.
For example, in the $Q_1 L_1 \bd_2$ direction given above,
the Yukawa couplings in the superpotential,
$\left(
\lambda_u H_u \bu_1 +
\lambda_d H_d \bd_1 +
\lambda_e H_e \be_1 +
\lambda_d H_d Q_2 \right) \langle \phi \rangle,$
lead to masses for quark, lepton and
Higgs superfields.
Gauge symmetries are also broken along flat directions,
with the broken gauge supermultiplets gaining mass
by the super Higgs mechanism,
$m_g = g \langle \phi \rangle$.
In the $Q_1 L_1 \bd_2$ example, the standard model gauge
group is broken to a $SU(2)_C \times U(1)$ subgroup.
Along more general flat directions the gauge group is
typically completely broken.

In the early universe the relevant scale for excitations
in a radiation dominated era is of course the temperature.
Analogously the scale for quantum deSitter fluctuations
in an inflationary era is the Hubble constant.
Far out along a flat direction the modes which
gain mass from Yukawa and gauge couplings
become heavier than these
excitation scales, and therefore decouple.
This is why the moduli space is the relevant subspace
on which the dynamics takes place when the
fields are large.

The directions referred to above as ``flat'' can
be lifted by supersymmetry breaking
and terms in the superpotential.
The resulting potential is of central importance to the discussion
of the evolution along flat directions in the early universe.
For the AD mechanism the origin of the potential terms which
violate the $U(1)$ carried by the direction is also crucial.
First consider the potential arising from the superpotential.
In general since a flat direction can be represented by
an invariant operator, it can appear to some power
in the superpotential.
These superpotential terms could vanish accidentally and not lift
the direction; the nonrenormalization theorem makes this
technically natural.
Certain symmetries can also forbid all such terms in the superpotential,
as discussed at the end of this section.
However to be as general as possible,
unless stated otherwise, we will assume all superpotential
terms consistent with gauge symmetry and $R$ parity are present.

The nonrenormalizable terms in the superpotential which lift
the flat directions are of two types.
First, since the direction can be written as an invariant,
$X$, it can appear to some power
\beq
W = \frac{\lambda}{nM^{n-3}} X^k = \frac{\lambda}{nM^{n-3}} \phi^n
\label{WX}
\eeq
where $X=\phi^m$, $n=mk$, and $M$ is some large mass scale
such as the GUT or Planck scale.
Under our assumptions above, the lowest value of $k$ is 1 or 2
depending on whether the direction is even or odd under $R$ parity.
The second type of term which lifts the flat direction contains
a single field not in the flat direction and
some number of fields which make up the direction,
\beq
W = \frac{\lambda}{M^{n-3}} \psi \phi^{n-1}
\label{Wpsi}
\eeq
For terms of this form, $F_{\psi}$ is nonzero along the flat direction.
An example of this type is the direction
$\bu_1 \bu_2 \bu_3 \be_1 \be_2$ which is lifted by
$W = (\lambda/M) \bu_1 \bu_2 \bd_2 \be_1$,
since
$F_{\bd_2}^*=(\lambda/M) \bu_1 \bu_2 \be_1$ is nonzero along
the direction.
In the flat space limit, with minimal kinetic terms,
the lowest order contributions of either
type of superpotential term, (\ref{WX}) or (\ref{Wpsi}),
give a potential
\beq
V(\phi) = \frac{|\lambda|^2}{M^{2n-6}} (\phi^* \phi)^{n-1}.
\label{genericpotential}
\eeq
 These terms always dominate the potential
for sufficiently large field value.    While the soft terms discussed below
can in principle have either sign,
these terms make a positive contribution
to the potential (provided $\phi \ll M_p$).
This has the consequence, as discussed in subsequent sections, of limiting
the fields to be parameterically less
than $M_p$.
All the flat directions listed in the Appendix can be lifted by
nonrenormalizable operators of the type discussed above with
$n \leq 6$.

The superpotential contribution to the potential (\ref{genericpotential})
has the interesting property that it conserves the $U(1)$ carried
by the flat direction despite the fact that
the superpotentials (\ref{WX}) and (\ref{Wpsi}) violate the $U(1)$.
This is because for a single term in the superpotential
there is always an accidental $R$ symmetry
under which $\phi$ has charge $R=2/n$.
Higher order operators can violate this accidental symmetry
in the potential
through interference with the lowest order term.
However, the coefficients of such terms are suppressed by additional
powers of the heavy scale $M$, and so are subdominant.
In principle, with multiple flat directions,
such interference terms could
arise at the same order as the $U(1)$ conserving terms.
This would require two flat directions made out of the
same number of fields, but with different $B-L$.
Examination of the list of flat directions in the
appendix, reveals that this occurs
for directions made of five fields.  In this case,
$\phi^5=LL\bd \bd \bd$ carries $B-L=-3$,
while all other directions made of five fields carry $B-L=1$.
However, for this example, including all superpotential terms
consistent with $R$ parity gives a $B-L$ conserving potential which
lifts the directions at lower order.
The $B-L$ violating interference terms are then again subdominant.
We therefore conclude that
nonzero $F$ terms arising from the nonrenormalizable superpotential
give rise (predominantly) to a $U(1)$ conserving potential
(\ref{genericpotential}).

The other source of potential terms for ``flat'' directions
is supersymmetry breaking.
In the flat space limit these can be represented
by soft breaking terms.
The general form of the soft terms is fixed.
The lowest order term is just a mass term
\beq
V(\phi)=m^2 \phi^* \phi
\label{softmass}
\eq
In addition if there are self couplings of the flat direction
in the Kahler potential or
superpotential (as discussed above), $A$ type terms can arise.
The lowest order $A$ terms are of the form
\beq
V(\phi)=\frac{A}{M^{n-3}} \phi^n
\label{softA}
\eq
where there are $n$ fields in the flat direction.
Assuming $R$-parity is unbroken in the early universe
all $A$ terms for standard model
flat directions are nonrenormalizable.
There are two important points to note about the
soft terms.
The first is the magnitude of  these terms.
In the present universe, assuming hidden sector supersymmetry
breaking, both $m$ and $A$ are
of order the weak scale, $\mgravitino$.
However, as we show in the next section, the
finite energy density in the early universe necessarily
breaks supersymmetry, inducing soft parameters
of order the Hubble constant,
$m \sim H$, and $A \sim H$ for $H > \mgravitino$.
This differs from the usual assumption that the soft
parameters are order $\mgravitino$ in the early universe,
and has dramatic consequences for the evolution.
The second important point is that the $A$ term violates
the $U(1)$ carried by the flat direction.
This will be the source of $B-L$ violation necessary
to generate a net $B-L$ in the evolution of
the AD flat direction.
In addition, the coefficient of the $A$ term is in general complex.
The relative phase between this and the ``initial'' phase of
the flat direction is the source of $CP$ violation
in the AD mechanism.
As discussed in section 4, all the potential terms
(\ref{genericpotential}), (\ref{softmass}), and (\ref{softA})
turn out to be important in determining the evolution
of the flat direction.

It is worth noting that because of the nonrenormalization theorem
for the superpotential, even operators consistent with
all symmetries need not appear in the superpotential.
In certain instances the absence of the gauge invariant
operators which could lift the flat direction can be
guaranteed by an $R$ symmetry.
Directions of this type are therefore
exactly flat in the supersymmetric limit.
Only soft terms contribute to the potential along such directions.
In the absence of a superpotential the potential
does not necessarily grow like a power for large fields.
Field values of order $M_p$ can therefore in principle
develop, and
higher order terms in the soft potential then become
important.
The general form of the $U(1)$ conserving soft potential
is (again, specializing for simplicity to the case of a single
field)
\beq
V(\phi)=m^2 M_p^2 ~f(\phi^* \phi/M_p^2)
\label{softpot}
\eq
where $f$ is some function.
Likewise the general $U(1)$ violating soft potential in this
case is
\beq
V(\phi)=m^2 M_p^2 ~g(\phi^n / M_p^n)
\label{softApot}
\eq
Notice that
the $U(1)$ violating terms start at order
$m^2$.
This is because $A$ terms (with coefficient $m$)
are proportional to $W$ and so vanish if $W=0$.
As shown in the next section the scale for $m$ again
turns out to be the Hubble constant for $H > \mgravitino$.
The scenario for the evolution of the AD field
with $W=0$ turns out
to quite different than in the case with a superpotential.
This is in fact the picture that was originally adopted
by Affleck and Dine \cite{ad}.
We briefly comment on the cosmology of the AD mechanism
for this case in section (\ref{flatscenario}).

The tree level superpotential also vanishes exactly for string
moduli.
The potential for these directions therefore also arises
from soft terms of the form (\ref{softpot}), again
with the soft parameters set by the Hubble constant
for $H > \mgravitino$.
We comment on the modification and possible solution
of the Polonyi problem
in the presence of such terms in section (\ref{stringevol}).

\section{Supersymmetry Breaking in the Early Universe}

As discussed in the previous section, flat directions can be
lifted by supersymmetry breaking and nonrenormalizable terms in the
superpotential.
In this section the potential along flat directions
arising from supersymmetry breaking in the early universe is   considered.
Generally it has been assumed that in the early universe
the relevant scale for the soft breaking
parameters $m$ and $A$ are of order $\mgravitino$ (assuming
hidden sector supersymmetry breaking).
Our main observation is that the finite energy density in the early
universe necessarily breaks supersymmetry.
As discussed below, for $H > \mgravitino$ this breaking
is dominant over breaking from a hidden sector,
and determines the soft potential along flat directions.

First consider how the finite energy density breaks supersymmetry.
A nonzero expectation value for the energy density,
and therefore the Hamiltonian, implies
the supercharge does not annihilate the vacuum, thereby
breaking supersymmetry.
The specific form of the breaking depends on the cosmological
epoch.
During an inflationary epoch the vacuum energy is positive by
definition.
Nonzero vacuum energy necessarily requires finite $F$ and/or
$D$ components for some matter fields, thereby signaling supersymmetry
breaking.
In a post inflationary epoch before reheating
occurs, the
energy density is dominated by the oscillations of the inflaton.
Again the time averaged vacuum energy is nonzero, thereby
breaking supersymmetry.
Supersymmetry is also broken in a radiation dominated era.
Here the boson and fermion thermal occupation numbers are distinct
so the background is not supersymmetric.
A similar quantum mechanical effect also exists during inflation;
deSitter fluctations give bosons and fermions distinct
nonzero occupation numbers.
As discussed below, these thermal and quantum effects
turn out to be less important
at large field values than the classical supersymmetry breaking from the
finite vacuum energy.

The finite energy supersymmetry breaking can be transmitted to
flat directions by either renormalizable or nonrenormalizable
interactions.
The effect of renormalizable interactions
is contained in the effective potential arising from integrating
out the states which gain a mass
along the direction,
$m_{\perp} = g \langle \phi \rangle$, where $g$ here represents
a gauge or Yukawa coupling.
In flat space at zero temperature supersymmetry guarantees that the
bosonic
and fermionic functional determinants
in the effective potential
cancel to all orders.
At finite temperature or in deSitter space, the
boson and fermion occupation numbers differ, and a nonvanishing
effective potential does arise.
However, at finite temperature for $\phi \gg T/g$,
the states which gain a mass along the flat direction effectively decouple,
$m_{\perp} \gg T$.
The effective potential arising from
integrating out these states is therefore exponentially
suppressed in this region, and can be neglected.
In deSitter space, where the Hubble constant sets the scale for
quantum excitations,
an analogous decoupling takes place
for $\phi \gg H/g$.
For large field values the induced potential from renormalizable
interactions is therefore unimportant \cite{yukawanote}.

Nonrenormalizable interactions can also transmit the supersymmetry
breaking to flat directions.
This contribution to the potential arises from integrating out fields
which do not gain a mass along the flat direction.
Unlike the case of renormalizable interactions,
this effective potential can induce a mass for a flat direction which
is roughly independent of the magnitude of the fields (so
long as it is less than $M_p$).
For large fields, nonrenormalizable interactions are therefore
more important than renormalizable ones.
To illustrate this effect consider the global limit with
a term in the Kahler potential of the form
\beq
\delta K =
\intD ~ {1 \over M_p^2}\chi^{\dagger} \chi \phi^{\dagger} \phi
\label{kahlercoupling}
\eq
where $\chi$ is a field which dominates the energy density of
the universe, $\phi$ is a canonically normalized flat
direction, and
$M_p = m_p/ \sqrt{8 \pi}$ is the reduced Planck mass.
No symmetry prevents such a term, which can be present already
at the Planck scale.
In fact, the existence of such operators is guaranteed
in the presence of Yukawa couplings since they are necessary
counterterms for operators
generated by loop diagrams \cite{maryk,bpr}.
If $\chi$ dominates the energy density, then
$\rho \simeq \langle \intD \chi^{\dagger} \chi \rangle$.
In a thermal phase the expectation value is just
the thermal mean value of the $\chi$ component kinetic terms,
$\rho \simeq g_* T^4$.
During inflation it is given by the inflaton
$F$ components,
$\rho = F^*_{\chi} F_{\chi} = V(\chi)$.
In the inflaton matter dominated era after inflation the expectation
value is again the total energy density,
$\rho = \dot\chi \dot\chi^* + V(\chi)$.
The interaction
(\ref{kahlercoupling}) therefore gives an effective mass for $\phi$ of
\beq
\delta {\cal L} = ( \rho / M_p^2) \phi^{\dagger} \phi
\eq
(note that a positive contribution in the Kahler potential gives
a negative contribution to $m^2$).
In a flat expanding background the energy density is related to the
expansion rate, $H$, by Einstein's equations,
$\rho = 3 H^2 M_p^2$.
This implies that the soft mass induced by the finite energy
supersymmetry breaking is $m^2 \sim H^2$.
This is a generic result, independent of what specifically dominates
the energy density.
For $H > \mgravitino$, this source for the soft mass is
more important than any hidden sector breaking.

In order to be concrete about the evolution along flat directions,
we will work under the assumption that there is
an inflationary phase sufficient to solve the horizon and flatness
problems. 
This requires $N > $ 60 $e$-foldings of the scale factor during
inflation.
We also assume this inflation gives rise to the density and
temperature fluctuations in the present universe.
In most models this occurs
for $H \sim 10^{13-14}$ GeV during inflation.
We also make the
standard assumption that after inflation, the universe enters
a matter dominated epoch in which the energy density is dominated by
coherent oscillations of the inflaton.  When the inflaton
decays, the universe enters an era in which the
energy density is dominated by the thermalized decay products
of the inflaton.  We assume that the associated ``reheat''
temperature, $T_R$,
is sufficiently low to avoid the ``gravitino problem" \cite{gravitino}.
The precise bound on $T_R$ depends on the gravitino
mass and branching ratios, but cannot be much larger than
$10^9$ GeV \cite{gravitino}.
The Hubble constant at reheating, $H_R$, and $T_R$ are related by
$g_* T_R^4 \sim 3 H_R^2 M_p^2$.
Avoiding the gravitino problem therefore requires $H_R \ll \mgravitino$.
With this restriction the induced potential
from finite density supersymmetry breaking
is only important (ignoring any pre-inflationary evolution)
during inflation and in the pre-reheating era dominated by
inflaton oscillations.
Since the inflaton dominates the energy density in both phases,
we only need to consider the soft potential induced by
couplings of
the inflaton to the flat directions.

Since the important couplings between the inflaton and
flat directions arise from Planck scale operators, supergravity
interactions should be included. The supergravity
scalar potential is
\beq
V = e^{K/M_p^2} \left(
D_i W K^{i \bj} D_{\bj}W^* - {3 \over M_p^2} |W|^2 \right)
+ {1 \over 8}f_{ab}^{-1} D^a D^b
\label{supergravity}
\eq
where $D_i W \equiv W_i + K_iW/M_p^2$ is the Kahler
derivative,
$W_i \equiv \partial W / \partial \varphi_i$,
$K^{i \bj} \equiv (K_{i \bj})^{-1}$,
$f_{ab}$ is the gauge kinetic function.
$W(\varphi)$ and $K(\varphi^{\dagger},\varphi)$
are the superpotential and Kahler potential,
$D^a \equiv K_{\varphi} T^a \varphi$,
where $\varphi$ includes in general the flat directions,
inflaton(s), and hidden sector.
By assumption the inflaton dominates the energy density
during inflation, and prior to reheating.
The largest piece of (\ref{supergravity}) is then for the
inflaton.
It is perhaps unlikely that $D$ terms in the inflaton sector
give a significant contribution to the inflaton
potential \cite{Dinf}.
The potential is not sufficiently flat along $D$ non-flat
directions to give a reasonable number of $e$-foldings
\cite{rt,moduliinfa,moduliinfb}.
If the inflaton potential arises from $F$ terms in some sector,
\beq
V(I)\simeq e^{K(I^{\dagger},I)/M_p^2}
\left( F_I^* F^I
-\frac{3}{M_p^2} |W(I)|^2 \right)
\label{fterminflation}
\eq
where
$ F_I^* F^I \equiv D_IW(I) K^{\bar{I}I} D_{\bar{I}}W^*(I^*)$.
The term in parenthesis necessarily has positive expectation value and
a nontrivial potential along flat directions is obtained as described
below.
Even if $D$ terms dominate the inflaton potential,
a nontrivial potential along flat directions can result.
In this case Kahler potential couplings
(such as (\ref{kahlercoupling}))
give a nontrivial potential along flat directions
from $K_I T^a I$, where $I$ is the inflaton.

Under the assumptions spelled out above,
the induced soft potential for the flat directions arises
from couplings to the inflaton $F$ components.
However,
the specific form of the induced soft potential can depend on
the scalar value of the inflaton.
During inflation $I \sim M_p$ is possible.
In fact in ``natural'' models $I$ changes by order $M_p$
per $e$-folding during inflation \cite{rt,moduliinfa,moduliinfb}.
In this case
since the inflaton vacuum energy is necessarily positive,
$W(I)/M_p$ can be at most the same order as $D_I W(I)$.
Using the relation between the energy density and Hubble
constant, $\rho = 3 H^2 M_p^2$,  gives the scales
$D_I W(I) \sim H M_p$ and $W(I) \sim H M_p^2$.
The pure supergravity corrections to the potential
are considerably simplified if
$I \ll M_p$.
In this limit $K_I \ll M_p$,
$D_I W \rightarrow W_I$, and
$|W|/M_p \ll W_I$.
The inflaton potential then reduces to
$V(I) \simeq W_I^* W^I $ (up to corrections of
${\cal O}(I/M_p)^2$).
$I$ may be of order $M_p$ during inflation, but after inflation
$I \ll M_p$
if, as we assume, the energy
density after inflation is dominated by the coherent
oscillations of the inflaton field.
In this era we assume that the relevant potential for the inflaton
is just that of a harmonic oscillator,
$V(I) \simeq m_I^2 (I-I_0)^*(I-I_0)$.
 If the inflaton potential is at all natural, it is unlikely that
the mass of the inflaton, $m_I$, is much smaller than
$H_I$, the Hubble constant during inflation.
Therefore well after inflation
$\langle I^* I \rangle   \sim
 (H^2 /m_I^2) M_p^2
\sim (H^2 / H_I^2) M_p^2 \ll M_p^2$
and the supergravity corrections are small.
This distinction will only be important for the induced $A$
and $\mu$ terms discussed below.

The general form for the induced potential from
(\ref{supergravity}) along flat directions depends on whether
or not the flat direction is lifted by nonrenormalizable
terms in the superpotential.
First consider terms which are independent of the superpotential
along the flat direction.
In this case the potential for a  flat direction
arising from supersymmetry breaking terms
in the supergravity potential comes from the following sources:

\noindent
1)  The $e^{K/M_p^2}$ prefactor
$$
e^{K(\phi^{\dagger}, \phi)/M_p^2} V(I)
$$
2) Cross terms in the Kahler derivative between the flat direction
Kahler potential and inflaton superpotential
$$
K_{\phi} K^{\phi \bphi} K_{\bphi}~ \frac{|W(I)|^2}{M_p^4}
$$
3) Kahler potential couplings between the inflaton and
flat direction
$$
K_{\phi} K^{\phi \bar{I}} D_{\bar{I}} W^*(I) \frac{W(I)}{M_p^2} ~ + h.c.
$$
With the scales for the inflaton potential potential
terms given above,
all these give the general form
\beq
V(\phi) = H^2 M_p^2 ~f (\phi / M_p)
\label{genericpot}
\eq
where $f$ is some function.
Notice that the overall scale of the
potential is set by the Hubble constant,
$V^{\prime \prime} \sim H^2$, and the scale for
variations in the potential is $M_p$.
This is the form of the induced potential for string moduli
or standard model directions which are exactly flat in
the supersymmetric limit.
 For string moduli the minimum of the induced potential
(\ref{genericpot}) is in general displaced by order $M_p$
from the true minimum arising from hidden sector supersymmetry
breaking.
However if there is a point of enhanced symmetry on the moduli
space, the potential (no matter what the source) is
necessarily an extremum about this point (for the potential
induced by the finite density breaking this follows since
the Kahler potential is a minimum about a symmetry point).
For standard model fields, the origin is always an enhanced
symmetry point, so the potential is always an extremum at the origin.

An important special case of the general form
(\ref{genericpot}) results for a minimal Kahler potential
for the flat direction,
$K(\phi^{\dagger},\phi) = \phi^{\dagger} \phi$.
Assuming $F$ terms dominate the inflaton energy density
as in (\ref{fterminflation}), and
using the relation between the energy density and expansion rate,
$V=3H^2 M_p^2$,
the resulting induced potential for $\phi \ll M_p$
is then just a mass term
$m_{\phi}^2 \phi^* \phi$, with
\beq
m_{\phi}^2 = \left(2 + \frac{F_I^*F_I}{V(I)} \right)
H^2  
\label{minimalmass}
\eq
For $I \ll M_p $, $V(I) \simeq F_I^* F_I$ as discussed above.
(These expressions should be corrected by
$V_F(I)/(V_F(I)+V_D(I))$ if inflaton $D$ components
contribute to the energy density).
The important feature of this contribution
is that it is {\it positive} with coefficient
of order one.
This will have important implications for the AD
mechanism discussed in the next section.
With minimal Kahler terms only, $\phi=0$ is stable
and the large expectation values required for baryogenesis
do not result.
With general Kahler terms though,
$|m_{\phi}^2| \sim H^2$, with either sign possible.
In this paper we will assume that this is the case.
However, it is possible to choose special forms for the Kahler
potential couplings between the flat direction and inflaton
which partially cancel the minimal supergravity induced
mass (\ref{minimalmass}).
It has been suggested that no-scale like forms of Kahler potentials
(which often arise at tree level in string theory) might
accomplish this \cite{stewart,gmo}.
It is also important to recognize that symmetries
can protect a compact subspace of a flat direction
from receiving a soft potential.
This is the case for a Goldstone boson of a spontaneously
broken symmetry.

There are additional contributions to the potential induced from
the finite density supersymmetry breaking if a flat direction is
lifted by nonrenormalizable terms in the superpotential.
These arise from:

\noindent
1) cross terms in the Kahler derivative between the
derivative of the flat
direction superpotential and inflaton superpotential,
$$
W_{\phi} K^{\phi \bphi} K_{\bphi} \frac{W^*(I)}{M_p^2} ~ + h.c.
$$
2) cross terms between the flat direction and inflaton superpotential
$$
\left(  \frac{1}{M_p^2} K_{I} K^{I \bar{I}} K_{I }  - 3 \right)
\left(  \frac{W(\phi)^* W(I)}{M_p^2} ~ + h.c. \right)
$$
3) Kahler potential couplings between the flat direction and
inflaton
$$
W_{\phi} K^{\phi \bar{I}} D_{\bar{I}} W^*(I) ~ + h.c.
$$
With the nonrenormalizable superpotential terms
(\ref{WX}) and (\ref{Wpsi}) and the inflaton scales given above,
all these have the form of a generalized $A$ term
\beq
V(\phi) = H M_p^3 g(\phi^n / M_p^n)
\eq
where $g$ is some function.
The induced $A$ terms have the important effect of defining the
intial phase of the AD field, as discussed in the next
section.
The possible $A$ terms considerably simplify if $I \ll M_p$ as
is the case after inflation.
Using the inflaton scales given above for $I \ll M_p$,
only the $W_{\phi} K^{\phi \bar{I}} W^*_{\bar{I}} ~ + h.c.$ term
can contribute significantly.
Since all the fields are $\ll M_p$ in this
case the Kahler potential couplings can be expanded in powers
of $M_p$.  The only term of the required form which gives rise
to an $A$ term is then
\beq
\frac{1}{M_p} \intD~ I \phi^{\dagger} \phi  
\eq
If $I$ is a composite field rather than an elementary singlet,
then only terms bilinear in the canonically normalized
inflaton field can appear in the Kahler potential and
such a term does not exist.
It is therefore possible in some models for the induced $A$ terms
to vanish after inflation.
The same conclusions hold for other dimension 3 soft terms and
the induced $\mu$ term as discussed in section \ref{nbssection}.

In sum, for $H > \mgravitino$,
the soft potential for the flat direction (away from the origin)
is set by the supersymmetry breaking due the finite energy density,
with soft breaking scale given by the Hubble constant. 
This is our most important result.
Previously it had been (implicitly) assumed that
the hidden sector supersymmetry breaking set the scale for the
soft potential.
Self couplings from nonrenormalizable superpotential terms
have also not been consistently included in discussions of the evolution
of flat directions.


\section{Evolution of the AD Scalar}

The evolution of the fields parameterizing a flat direction
is governed by the classical equations of motion.
For the AD mechanism, assuming the baryon number
does not average to zero over the current horizon size,
only the zero mode of the field is relevant.
The equation of motion for the zero mode is just that of
a damped oscillator
\beq
\ddot\phi + 3 H \dot\phi + V^{\prime}(\phi) = 0
\label{dampedosc}
\eq
where the damping term, proportional to the Hubble constant,
arises because of the expanding background.
The behavior of the solutions of (\ref{dampedosc}) are well know.
For $H^2 \gg V^{\prime \prime}(\phi)$ the field is overdamped and the
friction term dominates the evolution.
For $H^2 \ll V^{\prime \prime}(\phi)$ the field is underdamped and
the inertial term dominates the evolution.
Previously, the implicit assumption has been that the potential
for $\phi$ arose from hidden sector supersymmetry breaking,
$V^{\prime \prime}(\phi) \sim \mgravitino^2$.
If this were the case, at early times when $H \gg \mgravitino$
the field would be highly overdamped, and effectively frozen
at some ``initial'' value.
When $H \sim \mgravitino$ the field would begin to oscillate
about a local minimum.
However, as discussed in the previous section the scale for
the soft potential arising from the finite density supersymmetry breaking
is the Hubble constant.
At early times the fields are parameterically near critically
damped.
During inflation when $H$ is roughly constant, the fields
can therefore very effectively evolve to an instantaneous minimum.
This has very important consequences for the AD
mechanism of baryogenesis, which requires large field values to
develop.

Crucial in assessing the possibility of baryogenesis is the sign
of the induced mass squared at $\phi=0$.
As discussed in the previous section, with minimal Kahler terms
the $m^2$ is of order $H$ with positive coefficient.
In this case as long as the field is within the basin
of attraction (which is very likely for directions which
are lifted by nonrenormalizable terms)
the average value of the field evolves to
$\phi=0$ exponentially in time.
This is in contrast to the usual statement that ``scalars are not
damped during inflation.''
Within supergravity, scalars can be very effectively damped away
during inflation.
After inflation no coherent production results, and the AD
mechanism does not occur.
Quantum deSitter fluctuations do excite the field with
$\langle \delta \phi^2 \rangle \sim H^2$ for $m \sim H$,
but with a correlation
length of $l \sim {\cal O}(H^{-1})$.
Any resulting baryon number then averages to zero over the present
universe.
In addition, the relative magnitude of the $B$ violating $A$ term
in the potential is small for $H \ll M$.
One possibility might be that $m^2 >0$ but $m^2 \ll H^2$.
The correlation length for deSitter fluctuations in this
limit is $l \simeq H^{-1}e^{3H^2/2m^2}$ \cite{desitter}.
This
is only large compared to the horizon size if  $(m / H)^2 < \frac{1}{40}$.
Although baryogenesis may be possible in this case,
as discussed in section 3,
$m^2 \ll H^2$ requires arranging couplings in the Kahler
potential partially cancel the minimal supergravity contribution
(\ref{minimalmass})
to $m^2$ over a wide range of values for the inflaton field
(including after inflation).
Although this is possible in toy models, it seems likely to require
fine tuning in any realistic example.

However, if the sign of the induced mass squared is negative a large
expectation value for a flat direction can develop.
With nonminimal Kahler terms this is perhaps just as likely as
a positive mass squared.
The magnitude of the field is then set by a balance with
nonrenormalizable terms
in the superpotential which lift the flat direction.
The post inflationary evolution turns out to be remarkably
simple and independent of the details of the potential.
We first summarize the salient features of the evolution in the
negative mass squared scenario and then explain each point
in more detail in subsequent subsections.

\begi

\item During inflation, the AD field evolves exponentially
to the minimum of the potential, determined by the induced
negative mass squared and nonrenormalizable term in the superpotential.
This process may be thought of
as establishing ``initial conditions" for the subsequent
evolution of the field.
The $B$ violating $A$ terms play an important role
in determining the  initial
phase of the field.

\item Subsequent to inflation, the minimum of the potential
is time dependent (as it is tied to the instantaneous value of the
Hubble parameter). The AD field oscillates
near this time dependent minimum with decreasing
amplitude

\item When $H \sim \mgravitino$ the soft potential arising from
hidden sector supersymmetry breaking becomes important and the sign
of the mass squared becomes positive.
At this time, the $B$-violating $A$ term arising from the hidden
sector is of comparable importance to the mass term,
thereby imparting a substantial baryon number to the condensate.
The fractional baryon number carried by the condensate is near
maximal,
more or less independent of the details of the flat direction.  Subsequent
to this time, the baryon number violating operators are negligible
so the baryon number (in a comoving volume) is constant.

\item The inflaton  decays when $H < \mgravitino$ (consistent
with the gravitino bound on the reheat temperature).
The baryon to entropy ratio subsequent to this is
\beq
{n_b \over s} \approx
{n_b\over n_\phi}{T_R \over m_\phi}{\rho_\phi \over \rho_I}
\eeq
where $n_b$ and $n_{\phi}$ are baryon
and AD field number densities, $T_R$ is the reheat temperature,
$m_\phi \sim \mgravitino $ is the low energy mass for the AD field,
and $\rho_\phi$ and $\rho_I$ are the AD field
and inflaton mass densities (both at the time of inflaton decay).

\endi

\noindent
The final baryon density depends
principally on the reheat temperature and the dimension
of the operator which stabilizes
the flat direction (in the supersymmetric limit), through
the factor $\rho_\phi/\rho_I$. The net baryon number density
is a robust prediction which
depends only weakly on the other variables in the problem,
such as the numerical values of the coupling constants
(and their phases) appearing in the superpotential.

In the following four subsections we discuss details of the
evolution and resulting baryon to entropy ratio
in the negative mass squared scenario
outlined above, assuming
the flat direction is lifted by nonrenormalizable operators.
In section \ref{positivescenario} the evolution for positive
(small) mass squared is considered.
In section \ref{flatscenario} we consider the evolution in the
case that the direction is exactly flat in the supersymmetric limit.

\subsection{The Inflationary Epoch}

\label{infepoch}

The large Hubble scale mass is clearly important to the evolution
of the field, as the field is parameterically near critically
damped.
We now consider in detail the evolution of the AD field
during inflation.
The flat direction is assumed to be stabilized, even in the
absence of supersymmetry breaking, by a high dimension operator
in the superpotential of the form (\ref{WX}) or (\ref{Wpsi}).
During inflation the Hubble parameter is roughly constant.
Given the discussions of sections 2 and 3 the relevant
potential during inflation then takes the form
\beq
V(\phi) = -cH_I^2 |\phi|^2 +
\left( {a\lambda H_I \phi^n \over n M^{n-3}} + ~h.c.~ \right) +
|\lambda|^2{|\phi|^{2n-2} \over M^{2n-6}}
\label{veff}
\eeq
where $c$ and $a$ are constants of ${\cal O}(1)$, and
$M$ is some large mass scale such as the GUT or Planck
scale.
For $H_I \gg \mgravitino$ soft terms arising from the hidden
sector are of negligible importance.
For $c > 0$
the potential (\ref{veff}) has an unstable extremum at the origin.
As discussed at the beginning of section 3, there is a
contribution to the potential for $\phi \sim H_I$ coming
from deSitter fluctuations of the fields coupled to the flat
direction by renormalizable couplings.
This gives a positive mass squared contribution to the free energy
of $\delta m^2 \sim g^2 H_I^2$,
where $g$ is a gauge or Yukawa coupling.
However, for $c \sim {\cal O}(1)$ the origin remains unstable.
Even if the origin is a local minimum from this effect
(which might happen if a large number of fields become massless
at $\phi=0$)
the global minimum at large $\phi$ is unaffected.
For very large $\phi$, (\ref{veff}) grows as
$|\phi|^{2n-2}$.
For $H_I \ll M_p$ this limits $\phi \ll M_p$ just on energetic
grounds.
The only soft terms which are
important are therefore the lowest order ones,
namely the mass and $A$ terms.

The minimum of the potential (\ref{veff}), is given by
\beq
|\phi_0| = \( { \beta H_I M^{n-3} \over  \lambda }
\)^{1 \over n-2}
\label{phimin}
\eeq
where $\beta$ is a numerical constant which depends on
$a$, $c$, and $n$.
Notice that $\phi_0$ is parameterically between
$H_I$ and $M$.
For example, with $H_I \sim 10^{13}$ GeV,
$M / \lambda \sim M_p$, and $n=4$, $\phi_0 \sim 10^3 H_I$.
The minimum is   larger for greater    $n$.
The $A$ term in (\ref{veff})
violates the $U(1)$ carried by $\phi$ and gives $n$
discrete minima for the phase of $\phi$.
The potential in the angular direction goes like
$\cos(\theta_a + \theta_{\lambda} + n \theta)$ where
$\phi = |\phi|e^{i\theta}$, etc.
During inflation if $c$ is not too small, the field quickly settles
into one of the minima.

In order to see just how fast the field evolves to the minima,
it is useful to consider explicitly the evolution of the magnitude,
ignoring for the moment the $A$ term.
At the beginning of inflation $\phi$ might be arbitrary.
However a simple constraint arises by requiring that the
energy density of the inflaton, $3H_I^2 M_p^2$, be greater than that
in $\phi$
(otherwise inflation could not take place).
For $ M/ \lambda \sim M_p$ this  gives
$\phi / \phi_0 < (M_p/H_I)^{1/(n-1)(n-2)}$.
For $n=4$ this gives $\phi < 10 \phi_0$.
For larger $n$ the maximum value is even smaller.
As a worst case, suppose
$\phi$ did start near this maximal value.
Then we expect that the field oscillates rapidly
with a period
much less than $H_I^{-1}$.  The amplitude of the oscillations,
$\phi_m$, is expected to decrease with a characteristic
time $H_I^{-1}$.
The time rate of change of the energy in $\phi$ can be found
from the equations of motion,
\beq
{dE \over dt}=\dot{\phi}{d \over d\phi}(T+V)=-3H_I{\dot{\phi}}^2
 = -6H_I(E-V)
\eeq
where $T$ is the kinetic energy.
Using the expression for $V(\phi)$ for large $\phi$ and
averaging over a period gives
$\dot\phi_m \simeq -6H_I/(2n-1) \phi_m$.
We therefore conclude that in the large $\phi$ regime,
$\phi$ decreases exponentially towards smaller values,
\beq
\phi_m \simeq e^{-6H_It/(2n-1)}\phi_i
\eeq
where $\phi_i$ is the initial value of the field with
espect to the origin.
Thus after just a few $e$-foldings $\phi$ is near a minima.

Once near a minimaum, the field evolves like
a damped harmonic oscillator.
For the region in which the potential
is approximately harmonic,
the equation of motion, neglecting the $a$ term, is
\beq
\ddot{\phi}^{\prime}+2H\dot{\phi}^{\prime}-2(n-2)cH_I^2\phi^{\prime}=0
\eeq
where $\phi^{\prime}=\phi-\phi_0$.
So long as $c$ is not too small,
the system will quickly settle into a minimum.
 The field undergoes deSitter fluctuations about the minimum,
with amplitude $\langle \delta \phi^2 \rangle \sim H_I^2$.
This is a small perturbation in the radial mode since
$\delta |\phi| / \phi_0 \sim H_I/ \phi_0 \ll 1$, and has
a very small correlation length $l \sim H_I^{-1}$.
If $a$ is not too small, the angular mode also has a mass of
order $H_I$.
The fluctuations are then also a small perturbation in this mode,
$\delta \theta \sim H_I / \phi_0 \ll 1$, again with a small correlation
length.
At the end of inflation, over regions large compared to the
current horizon size, $\phi$ is
left with essentially a constant ``initial'' phase.
As discussed in section 3 it is possible in principle
that the finite density $A$ term is very small
during inflation (if the inflaton is composite and $I \ll M_p$).
In this case there is no potential for the phase of $\phi$.
The phase then undergoes a random walk from deSitter fluctuations.
But by the end of inflation, the correlation length for $\delta \theta$
is necessarily larger than the current horizon.
So again the present universe is left with an essentially constant
(random) ``initial'' phase \cite{phasenote}.

We conclude that at the end of inflation, the average value of the field
is at one of its minima, with a large expectation value (\ref{phimin}).
In addition the field has a definite value for its phase,
which is constant over scales large compared to the present horizon.
This amounts to the ``initial'' conditions for the subsequent
evolution.

\subsection{Post-Inflation: Inflaton Matter Dominated Era}

\label{postinflation}

After inflation the universe enters a matter era dominated by the
coherent oscillations of the inflaton.
During a matter era the Hubble constant is related to the expansion
time by $H=\frac{2}{3}t^{-1}$.
The equation of motion for $\phi$ is then
\beq
\ddot \phi + \frac{2}{t} \dot \phi + V^{\prime}(\phi) =0
\label{phiteq}
\eeq
where $V(\phi)$ is still given by (\ref{veff}),
though the dimensionless
constants $c$ and $a$ may be different, and $H$ is now
time dependent.
To simplify the analysis, we will neglect $a$ during this
phase of the evolution.
(If $a \neq 0$ and different from that during inflation
then the phase simply evolves to a different value.)
The most important feature of (\ref{phiteq}) is that the
minimum of the potential, $\phi_0$, now decreases with time.
Since the potential grows like a power law for large $\phi$,
one might guess that if the field starts out not too far
from the minimum at early times, it will closely
track the minimum.

Greater insight into the solutions of (\ref{phiteq}),
can be gained by
making changes of variables.
Since the minimum decreases as a power law in $t$ it is
useful to rescale time as
$$
z=\log t
$$
and define the dimensionless field $\chi$
with respect to the instantaneous minimum
$$
\phi=\chi \phi_0(t) = \chi
 \left( \frac{ \beta}{ \lambda} M^{n-3} e^{-z}
   \right)^{\frac{1}{n-2}}
$$
where $\beta=\sqrt{c^{\prime}/(n-1)}$ for $a=0$, and
$c^{\prime}= \frac{4}{9} c$.
The equation of motion in these rescaled variables is then
\beq
\ddot\chi +
\left( \frac{n-4}{n-2} \right) \dot\chi -
\left[ c^{\prime} + \frac{n-3}{ (n-2)^2 } \right] \chi +
    c^{\prime} \chi^{2n-3} = 0 .
\label{rescaledeom}
\eeq
The rescaled problem is so simple because the
effective mass term, Hubble damping term, and acceleration
term are all homogeneous in $z$.
The equation of motion (\ref{rescaledeom}) has
two important properties.
First, there is a fixed point at
\beq
\bar{\chi} = \left( 1 +  \frac{n-3}{ c^{\prime} (n-2)^2 }
 \right)^{\frac{1}{2n-4}}
\label{fixedpoint}
\eeq
For reasonable values of the parameters this is just
slightly larger than the position of the instantaneous
minimum.
So if $\phi$ starts at this fixed point it remains there,
i.e. $\phi(t)=\bar{\chi}\phi_0(t)$, and $\phi$ tracks
just behind the decreasing minimum.
Second, the damping in the rescaled problem depends on $n$.
For $n > 4$ the effective damping in the rescaled
problem is positive.
In this case the fixed point is attracting.
For general initial conditions, the field oscillates
about $\bar{\chi}$ with decreasing amplitude.
For $n=4$ there is no damping of $\chi$.
The field $\phi$ therefore oscillates about the attracting point
with an envelope which decreases in time in proportion
to the instantaneous minimum.
For $n<4$ the damping is negative, but this corresponds to
a direction which is not even flat at the renormalizable level.
We conclude that for $n \geq 4$ the magnitude of $\phi$
decreases with the instantaneous minimum.


The oscillatory motion about the fixed point in the rescaled
problem is physically reasonable.
If $\phi$ starts at a large value it is underdamped
($V^{\prime \prime} \gg H^2$)
and gets driven to smaller values by the acceleration term.
Eventually it reaches small values of $\phi$ where
it is overdamped ($V^{\prime \prime} \ll H^2$) and slows
due to the friction term.
Still later,  the instantaneous minimum catches up and overtakes
$\phi$, again leaving it in an underdamped regime, and so forth.
As the instantaneous minimum decreases,
the field therefore naturally oscillates about a point at which
$V^{\prime \prime}(\phi) \sim H^2$, which is necessarily close
to $\phi_0(t)$.
Numerical evolution of (\ref{rescaledeom})
supports this picture, and the $n$ dependence explained above.

It is possible that the field does not start near the minimum
when the AD field potential during and subsequent to inflation
are very different. This would only occur when $I$ during
inflation is of order $M_p$, so that there can be terms
which change the qualitative form of the potential during
inflation but are negligible afterwards. In this case, one
can proceed using the adiabatic approximation of the
previous section. Here, because the Hubble constant
is time dependent, the field is only damped with a power
law dependence, and decreases at a rate
$\phi\propto t^{-1/(n-2)} $.

\subsection{Late Stage of Evolution: $H \sim m_{3/2}$}

The most interesting behavior of the fields is for
$H \sim m_{3/2}$.
Until this time, the quadratic term,
and $A$ terms are of comparable importance (unless $a \ll 1$),
and there is no sense in which baryon number is conserved.
Once $H \ll m_{3/2}$, the baryon number per comoving volume
is frozen.
The potential, including now the low energy soft terms
arising from hidden sector supersymmetry breaking is
\beq
V(\phi) = m_{\phi}^2 |\phi|^2
 - { \frac{c^{\prime}}{t^2}} |\phi|^2
+ \left( {(A m_{3/2} + aH) \lambda \phi^n \over n M^{n-3}} ~+h.c.~ \right)
 + |\lambda|^2 { |\phi|^{2n-2} \over M^{2n-6}}
\label{vtot}
\eeq
where $m_{\phi} \sim \mgravitino$.
At early times the field tracks near the
time dependent minimum as discussed in the last section.
Therefore when $H \sim \mgravitino$ all the terms in (\ref{vtot})
have comparable magnitudes.
Since the soft terms have magnitudes fixed by $\mgravitino$
the field is no longer near critically damped, but becomes
underdamped as $H$ decreases beyond $\mgravitino$.
In addition, the $\mphi^2 |\phi|^2$ term comes to dominate
the $-cH^2 \phi^2$ term as $H$ decreases.
The field therefore begins to oscillate freely
about $\phi=0$
when $H \sim \mgravitino$, with ``initial'' condition given
by $\phi_0(t)$ (eq. (\ref{phimin})) with $t \sim \mgravitino^{-1}$.
The oscillation of the field is the coherent condensate,
$\nphi \simeq \mphi |\phi|^2$.

Crucial for the generation of a baryon asymmetry are the $B$ violating
$A$ terms in (\ref{vtot}).
However, as discussed above when $H \sim \mgravitino$ all the
terms have comparable magnitude, including the $A$ terms.
Since $V_B \sim V_{\not B}$ when the field begins to oscillate freely
a large fractional baryon number is generated in the ``initial''
motion of the field when $m^2$ becomes positive.
Notice that in this negative mass squared scenario $n_b / \nphi$
is roughly {\it independent} of $\lambda / M$.
This is because the value of the field is determined precisely
by a balance of (negative) soft mass squared term and
nonrenormalizable supersymmetric term.
That the $B$ violating $A$ term also has the same magnitude
follows from supersymmetry since its magnitude is the root mean square
of the soft mass term and nonrenormalizable
supersymmetric term.
In this scenario there is no need for ad hoc assumptions about
the initial value of the field when it begins to oscillate
freely.
The expectation that $n_b / \nphi \sim {\cal O}(1)$ falls
out naturally.

The important role of $CP$ violation is also dictated
by the $A$ terms.
As discussed in section \ref{infepoch} at early times the
potential for the phase of $\phi$ goes like
$\cos(\theta_a + \theta_{\lambda} + n \theta)$.
As $H$ decreases below $\mgravitino$ the low energy $A$ term
becomes more important and the angular potential goes like
$\cos(\theta_A + \theta_{\lambda} + n \theta)$.
When the field begins to oscillate freely a nonzero
$\dot\theta$ is therefore generated if $\theta_a \neq \theta_A$.
This is of course required in order to generate a nonzero baryon number
since $n_b = 2 | \phi |^2 \dot\theta$.
The resulting baryon number therefore depends on the $CP$ violating
phase $\theta_a - \theta_A$, i.e. on
a relative phase between the inflaton and hidden sectors.
Alternately, as discussed in section (\ref{infepoch}) it is possible in
principle for $a$ to vanish during and after inflation.
The initial phase is then random (but constant
over scales large compared to the present horizon).
The $CP$ violation from the initial phase is then
effectively spontaneous.
It is interesting to note that if this is the case, the net
baryon number averaged over all inflationary domains vanishes.

Let us now consider in detail the numerical evolution of
the field equation in
this late stage of evolution.
In the discussion which follows, we will assume $a=0$ after
inflation, and take the initial phase, $\theta_i$,
 of $\phi$ as an input.
We have also done the analysis with $a \neq 0$ and find no
qualitative difference.
It is useful to once again work with rescaled variables.
The field is rescaled as
$$
\phi \rightarrow\left( \frac{ \mgravitino M^{n-3} }{ \lambda }
   \right)^{\frac{1}{n-2}}\phi
$$
{}From the arguments above and (\ref{phimin}), up to a numerical constant
of order unity, this is just the value of the field when
$H \sim \mgravitino$.
All other mass scales and time are rescaled with
respect to $\mgravitino$.
The equation of motion (\ref{vtot}) with $a=0$ and
$\theta_A + \theta_{\lambda} =0$
is then
\beq
\ddot{\phi} + \frac{2}{t} \dot{\phi}
 + \left( \mphi^2 - \frac{c^{\prime}}{t^2} \right) \phi
 + A \left( \phi^* \right)^{n-1}
 + (n-1) \left( \phi^* \phi \right)^{n-2} \phi
   = 0
\eq
The equation of motion for the real and imaginary parts
(appropriate for numerical integration) are
\begin{eqnarray}
 & &   \ddot\phi_R
   + \frac{2}{t} \dot{\phi}_R
   + \left( \mphi^2 - \frac{c^{\prime}}{t^2} \right)  \phi_R
   + A |\phi|^{n-1} \cos \left( (n-1) \theta \right)
   + (n-1) |\phi|^{2n-4} \phi_R
   = 0
     \nonumber \\
 & &   \ddot\phi_I
   + \frac{2}{t} \dot{\phi}_I
   + \left( \mphi^2 - \frac{c^{\prime}}{t^2} \right) \phi_I
   - A |\phi|^{n-1} \sin \left( (n-1) \theta \right)
   + (n-1) |\phi|^{2n-4} \phi_I
   = 0
    \nonumber \\
\label{rieom}
\end{eqnarray}
where $\phi = \phi_R + i \phi_I$, and
$\theta = {\rm Arg}~\phi$.
The initial $\phi$ and $\dot{\phi}$ for $t \ll 1$ were chosen
such that the field tracks the fixed point (ignoring the
$\mgravitino^2$ mass term).
The equations of motion (\ref{rieom}) were then integrated forward
in time to $t \gg 1$.
In this regime $n_b / n_{\phi}$ asymptotes to a constant value.
A typical trajectory in the $\phi$ plane is shown in
fig. 1 for $\mphi = c^{\prime} = -a = 1 $, $n=4$, and
$n \theta_i = \frac{9}{10} \pi$.
The field tracks near the fixed point until the
$\mgravitino$ mass and $A$ terms become important.
When $t \sim 1$ ($H \sim \mgravitino$) the field feels a
``torque'' from the $A$ term, and spirals inward in the
harmonic potential.
The nonzero $\dot{\theta}$ in the trajectory gives rise to
the baryon number.
For the trajectory in fig. 1 $n_b/n_{\phi} \simeq .8$, as
can be estimated by eye from the eccentricity of the ellipse.
The fractional baryon number carried by the condensate is
shown in fig. 2. as a function of $n \theta_i$ for
$\mphi = c^{\prime} = -a = 1$, and $n=4$.
Notice that $n_b/n_{\phi}$ is 0 for $\theta_i=0$ which corresponds
to a minimum of $V(\theta)$, and changes sign at $n \theta_i = \pi$,
which is a maximum of $V(\theta)$.
For this choice of parameters,
before spiraling in the harmonic part of the potential,
the field goes through one angular oscillation about the minimum of
$V(\theta)$ while the $A$ term is still important.
This is the origin of the zero in $n_b/n_{\phi}$
for $n \theta_i \simeq 0.65 \pi$.
For this initial value the field receives an equal and opposite
integrated torque as it oscillates through $V(\theta)$ while
the $A$ term is important.
Integrating over $\theta_i$, the rms $n_b/n_{\phi}$ is $0.3$.
This is typical of the result for other $n$ and reasonable
soft parameters.

At very late stages of the evolution when $H \ll m_{3/2}$,
the only potential term which is relevant in
(\ref{vtot}) is the soft mass term $\mphi^2 |\phi|^2$
which is of course $B$ conserving.
The baryon number created during the epoch $H \sim \mgravitino$
is therefore conserved by
the classical evolution of $\phi$ for $H \ll \mgravitino$.

\subsection{Baryon to Entropy Ratio}

\label{nbssection}

As discussed in the previous subsection, the fractional baryon
number stored in the condensate is likely to be near maximal,
independent of the order at which the flat direction is lifted.
The relevant physical quantity of interest however  is the baryon to
entropy ratio, which is $n_b/s \sim 10^{-10}$ in the present
universe.
In this section we show that $n_b/s$ depends in an essential way
only on the reheat temperature after inflation, $T_R$, and
the magnitude of the nonrenormalizable operator which lifts the flat
direction.  These in turn determine the fractional energy density
stored in the AD field.
The reheat temperature depends on details of the inflationary
model, and introduces some uncertainty in the final answer.
The dependence on some nonrenormalizable $B$ or $L$ violating
operator could in principle relate $n_b/s$ to some $B$ or $L$
violating process observable in the laboratory.
Unfortunately this is generally not the case because the $B$ or $L$
violation occur through higher dimension operators, the effects of which are
negligible at small field value. However, the
success of the preferred scenario is related to the lightest
neutrino mass, as discussed below.

Although $n_b / \nphi$ is not small, the total density in the condensate
$\rhophi \sim \mgravitino^2 \phi^2$,
is much smaller than the total density for $\phi_0 \ll M_p$.
For $H \sim \mgravitino$ the coherent oscillations of the inflaton
still dominate the energy density as discussed previously,
$\rho_I \sim 3 H^2 M_p^2$.
Using the estimate (\ref{phimin}) for $\phi_0$ at $H \sim \mgravitino$,
the fractional energy in the AD condensate at this time is
\beq
{\rho_{\phi} \over \rho_I} \approx \left (
{ \mgravitino M^{n-3} \over \lambda M_p^{n-2} }
   \right )^{2/(n-2)}.
\eeq
For $n=4$,
$ \rho_{\phi} / \rho_I \sim 10^{-16} (M / \lambda M_p)$,
while for $n=6$,
$ \rho_{\phi} / \rho_I \sim 10^{-8} (M^3 / \lambda M_p^3)^{1/2}$,
Notice for smaller $(\lambda/M^{n-3})$
the direction is effectively flatter, and $\phi_0$ and $\rho_{\phi}$
are larger.
A greater total energy is therefore stored in the oscillating condensate
for smaller $\lambda$ or larger $n$.
As discussed in section 3 the inflaton decays when $H < \mgravitino$.
Until the inflaton decays $\rhophi / \rho_I$ stays roughly constant
as both the AD condensate and inflaton redshift like matter.
The number density in the AD condensate is
$n_{\phi} = \rhophi / m_{\phi}$.
After the inflaton decays the baryon to entropy ratio is
therefore
\beq
{n_b \over {s}} \approx {n_b \over n_{\phi}}{T_R \over
m_{\phi}}{\rho_{\phi} \over \rho_I}.
\label{nbs}
\eeq
where $s \approx \rho_I/T_R$.
This formula only applies if $n_{\phi} < \rho_I / T_R$
at the time of decay.  This is well satisfied for $n=4$ or $6$.
If this inequality is not satisfied, the entropy is actually dominated
by the AD decay, and $n_b/s$ is order unity in this case.
For $T_R$ above the weak scale anomalous sphaleron processes
are in equilibrium.
So only directions with nonzero $B-L$ give a significant
baryon number in this scenario.
As long as the AD condensate decays
through $B-L$ conserving decays
after the inflaton, the
estimate (\ref{nbs}) is insensitive to the details of the decay.
Once the amplitude of the field becomes small enough for
degrees of freedom coupled to the flat direction by renormalizable
couplings to be excited by the thermal plasma
(i.e. $m_{\perp}=g \langle \phi \rangle < T$ where $g$ is a gauge
or Yukawa coupling), the condensate
can decay by $B-L$ conserving thermal scatterings.
The rate for this scattering is set by $T$ rather than $\mgravitino$
(as would be the case for a free decay \cite{ad}). 
For typical parameters the condensate evaporates
by thermal scattering some time after the inflaton decays.

In order for the estimate given above to apply,
the field must begin to oscillate freely
when $H \sim \mgravitino$, and evaporate by thermal scatterings
with the plasma at a later epoch.
Even though most of the inflaton energy is not converted to
radiation until $H \sim H_R < \mgravitino$,
subsequent to inflation but before reheating
there is still a dilute plasma
with temperature
$T \sim (T_R^2 H M_p)^{1/4}$
arising from the inflaton decay
products \cite{kt}.
Scatterings with this ambient plasma must therefore
be unimportant when $H \sim \mgravitino$.
This is the case if $g  \langle \phi \rangle > T$.
If this were not the case, the condensate would be highly
damped by thermal scatterings and potentially evaporate {\it before}
the epoch at which the baryon asymmetry is established.
Using the scaling for the temperature given above,
\beq
\left. { \langle \phi \rangle \over T} \right|_{H \sim \mgravitino}
\sim {1 \over \sqrt{T_R M_I}}
\left( { H M^{n-3} \over \lambda} \right)^{1/(n-2)}
\eq
For $n=4$,
$\phi/T \sim (M_I/T_R)^{1/2} (M/ \lambda M_p)^{1/2}$,
where $M_I = \sqrt{\mgravitino M_p}$,
while for $n=6$,
$\phi / T \sim (M_p/T_R)^{1/2} (M^3 / \lambda M_p^3)^{1/4}$.
It is clear that for the $n=6$ directions and
$ T_R < 10^9$ GeV,
the condensate survives intact when $H \sim \mgravitino$,
allowing for the successful
creation of baryon number.
However, the $n=4$ case is somewhat borderline, depending on
the value of Yukawa couplings along the direction,
and the scale of the nonrenormalizable operator
which sets the expectation value.
For example, if $M / \lambda \sim 10^3 M_p$,
thermal up quarks, with Yukawa coupling
of order $10^{-4}$, could scatter with the $L_i H_u$ directions
(for any $i$) unless $T_R < 10^6$ GeV.

The total density in the AD condensate, and therefore $n_b/s$,
is very sensitive to $n$,
the order at which the flat direction is lifted.
For $n >4$ with $M \sim M_p$ and a reasonable $T_R$, $n_b/s$
is generally too large, without additional entropy releases.
For example, $n=6$ naturally gives the correct $n_b/s$ only when
$T_R$ is of order the weak scale.
Such low reheat temperatures can in fact arise for composite
flat directions which act as
inflatons \cite{moduliinfa}.
However, for $n=4$
\beq
\frac{n_b}{s} \sim 10^{-10} \left( \frac{T_R}{10^6 ~{\rm GeV}} \right)
  \left( \frac{ 10^{-3} M}{\lambda M_p} \right)
\label{nbsfour}
\eq
The parameters in (\ref{nbsfour}) represent ones which satisfy the
constraint on $\phi/T$ to avoid the thermal scatterings discussed above.
This is quite a reasonable range for $T_R$ to be consistent with
the bounds from thermal gravitino production.
The estimate (\ref{nbsfour}) is in contrast the standard
scenario \cite{ad} in which $n_b/s$ is generally quite large
(as discussed in section \ref{flatscenario}).
The only directions which carry $B-L$ and can be lifted at $n=4$ in
the standard model are the $LH_u$ directions.
The nonrenormalizable operator is then
\beq
W = \frac{\lambda}{M} (LH_u)^2
\label{LHLH}
\eq
This operator may be present directly at the Planck scale, or
could be generated, as in $SO(10)$ GUTs, by integrating
out heavy standard model singlets, $N$, with couplings
$gLH_uN$, and Dirac masses $M_NNN$, giving $\lambda / M = g^2/m_N$,
where $g$ is a Yukawa coupling which could be less than
${\cal O}(1)$ \cite{LHnote}.
At low energies this is the operator which gives rise to
neutrino masses.
For baryogenesis along the $LH_u$ direction in this scenario,
$n_b/s$ can therefore be related to the
{\it lightest} neutrino mass since the field moves out furthest
along the eigenvector of $L_i L_j$ corresponding to the
smallest eigenvalue of the neutrino mass matrix.
\beq
\frac{n_b}{s} \sim 10^{-10} \left( \frac{T_R}{10^6~{\rm GeV}} \right)
 \left( \frac{10^{-8}~{\rm eV}}{m_{\nu}} \right)
\eq
Note that $n_b/s$ is inversely proportional to the neutrino mass.
A {\it smaller} nonrenormalizable term leads to a { \it larger}
baryon number \cite{limitnote}.
Assuming $T_R < 10^9$ GeV in order to satisfy the gravitino
bound \cite{gravitino}, then requires that at least one neutrino be
lighter than roughly $10^{-5}$ eV.
Including the considerations about thermal scatterings with the
condensate at $H \sim \mgravitino$ discussed above, would
reduce the allowed $T_R$ to roughly $10^6$ Gev, and therefore the upper
limit on the lightest neutrino to $10^{-8}$ eV.

It is intriguing that the $LH_u$ direction is so successful.
There are a few things worth noting about this particular
direction.
We haven't addressed at all the likelihood that any particular
flat direction is favored.
$LH_u$ is special in that it contains a Higgs field.
So for one of the $LH_u$ nonzero all the directions listed
in the appendix are lifted at the renormalizable level
except $LL \be$ and $LL \bd \bd \bd$.
Including nonrenormalizable operators, it is not hard to see that
these remaining directions will not have amplitudes larger
than that along $LH_u$.
So the estimate for $n_b/s$ is robust for this direction.

The $H_u$ content is also special for another reason.
Even with minimal Kahler terms at the high scale,
the induced mass squared can become negative after including
quantum corrections
(this is the origin of electroweak symmetry breaking in the radiative
breaking scenario).
The leading log correction just comes from renormalization
group evolution from the high scale down to the particle mass
which is being integrated out.
For a flat direction the running comes from
integrating out particles which couple
through renormalizable couplings.
So this amounts to running down to the scale
$Q \simeq \phi$, since the modes coupled to the flat direction
have a mass $m_{\perp}=g\phi$.
For the $LH_u$ direction
the $H_u$ component receives the largest modification
because of the large top quark Yukawa coupling.
With minimal particle content the one loop beta function for $m_{H_u}^2$ is
\beq
\frac{d}{dq} m_{H_u}^2 \simeq 3 g_2^2 m_2^2 + g_1^2 m_1^2
 - 3 \lambda_t^2 \left( m_t^2 + m_{\bar{t}}^2 + m_{H_u}^2 + A_t^2
   \right)
\label{betafcn}
\eq
where $q = - (4 \pi)^{-2} \ln(Q^2)$, $m_1$ and $m_2$ are the
hypercharge and weak gaugino masses,
and here $A_t \equiv A \mgravitino + a H$.
In principle all the mass parameters in (\ref{betafcn})
could be of order $H$ at early times.
The top Yukawa gives a negative contribution to the mass squared.
Now the physical mass squared for the flat direction is
$m^2_{LH_u} = m_{H_u}^2 + m_{L}^2 + |\mu|^2$, where the
last contribution comes from the effective $\mu H_u H_d$
term in the superpotential.
For $H > \mgravitino$ it is possible that the $\mu$ term
induced by the finite density is much smaller than $H$.
As discussed in section 3, after inflation $I \ll M_p$.
The only source for an induced $\mu \sim H$ term in this
era (ignoring any superpotential couplings) is therefore
\begin{equation}
{1 \over M_p} \int d^4 \theta ~ I H_u H_d
\end{equation}
If the inflaton transforms nontrivially under some symmetry
then such linear terms don't arise, and $\mu \ll H$
(likewise, $A$ terms and gaugino masses are not induced
with scales or order $H$ after inflation in the absence of
terms linear in the inflaton).
If in fact $\mu \ll H$ after inflation,
then it is possible that
$m_{LH_u}^2$ turns negative
at a high scale during and/or after inflation, allowing large
a large expectation to develop.
This is especially true in GUT models where the larger
representations can give even larger negative contributions to the
beta function between the Planck and GUT scales.
When $H \sim \mgravitino$, $m_{LH_u}^2$ can become positive from
the positive $\mu^2 \sim \mgravitino^2$ contribution.
Even though
$m_{H_u}^2 + \mu^2$ must turn negative from running
to the weak scale in order to drive electroweak symmetry breaking
in the present universe,
$m_{LH_u}^2$ can remain positive at late times
because of the $m_L^2$ contribution.
Whether this scenario for inducing a negative mass squared
actually is realized depends crucially on the beta
function and therefore particle content at very high scales,
and on the nature and couplings of the inflaton.
This scenario does in fact work for certain values
of standard model parameters.\footnote[1]{We thank
Diego Castano for verifying this scenario numerically.}
However, it might be that with minimal kinetic terms,
the $LH_u$ direction is the most likely
of all the directions which carry $B-L \neq 0$
to have a negative mass squared at early times.

\subsection{Evolution for Positive Mass Squared}

\label{positivescenario}

It is possible for the induced
mass squared along a flat direction to be either positive
or negative.
As mentioned at the beginning of section 4, baryogenesis
along a flat direction with $m^2 \ll H^2$
but $m^2 > 0$ may be possible if the induced mass
term is small.
If $m^2 \ll H^2$, large deSitter fluctuations can result
during inflation.
The correlation length for these fluctuations is
$l \simeq e^{3H^2/2m^2}$ \cite{desitter}.
The magnitude and phase of the field is correlated over this
scale.
If the baryon asymmetry is to be constant over the current
horizon scale, $(m/H)^2 < \frac{1}{40}$.
Ignoring any higher order terms in the potential, the fluctuations
reach an equilibrium distribution
after $N > (H/m)^2$ $e$-foldings with
$\langle \delta \phi^2 \rangle \simeq 3H^4/8 \pi^2 m^2$
\cite{desitter}.
Including higher order terms, the
distribution of fluctuations saturates at
$V(\phi) \sim H^4$.
However, large correlation lengths only arise
in regions where $\phi$ is highly overdamped, i.e.
$V^{\prime \prime}(\phi) \ll H^2$.
These deSitter fluctuations have been suggested as a possible
mechanism to obtain large ``initial'' field values for
baryogenesis \cite{adfluct}.
It is important to note though, that the induced mass squared
must be tuned to be small over the entire range of the inflaton
during inflation.

The induced mass must also be tuned to be numerically small
after inflation.
In a quadratic potential, the field evolves in the inflaton
matter dominated era following inflation with
$m^2 = c^{\prime}/t^2$, as
\beq
\phi(t) \simeq \left\{  \begin{array}{ll}
 \phi_i t^{-\frac{1}{2}} \cos
     \left( \frac{1}{2} \sqrt{4c^{\prime}-1} \ln(t/t_i) \right)
   &  ~~c^{\prime} \geq \frac{1}{4} \\
 \phi_i t^{-c^{\prime}}
   &  ~~c^{\prime} \ll 1
  \end{array}
  \right.
\eq
where $\phi(t_i)=\phi_i$
If the induced mass is not also small after inflation
the envelope of the field decreases rapidly as
a power law to small values.
For $c^{\prime} \geq \frac{1}{4}$, the envelope
of the field scales as $|\phi| \sim |\phi_i| \sqrt{H/H_{inf}}$.
However, if $c^{\prime} \ll 1$ after inflation,
ignoring for the moment higher order terms in the potential,
the field is highly overdamped
in regions where the correlation length is large,
$V(\phi)^{\prime \prime} \ll H^2$,
and therefore roughly consant after inflation.
Eventually, as the induced mass contribution in the potential
becomes small as $H$ decreases, the higher order terms in the
potential become more important.
Once the higher order terms become important,
$V(\phi)^{\prime \prime} \sim H^2$, and the field becomes
near critically damped and begins to move.
This evolution can be analyzed with rescalings
similar to those given in section 4.2.
For $c^{\prime} \ll 1$ there is again an attracting point at
\begin{equation}
\bar{\phi}(t) \simeq \left( {\gamma M^{n-3} \over  \lambda t}
  \right)^{\frac{1}{n-2}}
\end{equation}
where $\gamma = \sqrt{ (n-3)/(n-1)} / (n-2)$.
As the Hubble constant decreases the $\phi$ field therefore
oscillates about this value, which is very similar to the
attracting point in the negative mass squared scenario.
When the field begins to oscillate freely at $H \sim \mgravitino$
all the terms in the potential have roughly the same
magnitude, and near maximal fractional baryon number can result.
So the entirely (small) positive mass squared scenario is
parametrically the same as the negative mass squared case, and
gives a similar result for the baryon asymmetry.

\subsection{Evolution for W=0}

\label{flatscenario}

As discussed in section 2, it is possible that
the superpotential vanishes along a flat direction.
This can be enforced by a discrete $R$ symmetry.
This case may be thought of as the
$n \rightarrow \infty$ or $\lambda \rightarrow 0$ limit of the
preceding discussions.
For exactly flat directions
the potential arises solely from Kahler potential couplings, and
is of the form (\ref{softpot}) and (\ref{softApot}).
The typical scale for variations in the potential is therefore $M_p$.
If $\phi=0$ is unstable during and after inflation, a minimum
can occur for $\phi \sim M_p$ at early times.
When $H < \mgravitino$ the minimum at large $\phi$
from the induced soft potential must disappear.
There must be a minimum in the soft potential at $\phi=0$
from hidden sector supersymmetry breaking since
for MSSM fields $\phi \ll M_p$ in the present universe.
The initial value for the field when $H \sim \mgravitino$ is therefore
of order $M_p$.
The $B$ or $L$ violating
soft potential (\ref{softApot})
(arising from the Kahler potential couplings
or hidden sector supersymmetry breaking) is then roughly equal in
magnitude
to the conserving potential when the field begins to oscillate
freely.
A large fractional baryon number is stored in the condensate as in the
previous scenario,  and
$n_b/ \nphi \sim {\cal O}(1)$.
Now however, for $\phi \sim M_p$, the energy density stored in the
Affleck-Dine condensate is of order the energy density of the universe.
So it is no longer true that $\phi$ represents a small fraction of the
energy.
This is the situation originally considered by Affleck and Dine \cite{ad}.

Once the inflaton decays, there is a thermal background which can
in principle scatter off the condensate as in the previous scenario.
However, before inflaton decay the AD and inflaton fields
have roughly equal energy density.
The value of the AD field at the era of inflaton decay is
therefore $\phi \sim (\sqrt{g_*} T_R / \mgravitino) T_R \gg T_R$.
So any fields which have renormalizable couplings to $\phi$
(and therefore gain a mass of $m_{\perp}=g \phi$) are too heavy to be
excited by the thermal plasma.
As a result, the condensate remains after inflaton decay, and
comes to dominates the
energy density since the plasma energy density redshifts away.
Once $m_{\perp} < m_{\phi}$, decays through renormalizable
couplings become kinematically accessible.
For $g > 10^{-5}$ the condensate decays essentially
as soon as such decays are allowed,
with an effective reheat temperature
$T_{R,\phi} \sim m_{\phi} / \sqrt{g}$ \cite{rt}.
Since the condensate decay is itself the dominant source of entropy,
$n_b/s$ is much different than in the previous scenario.
For $T_{R,\phi} \gg m_{\phi}$ the decays give a relatively large
number of low energy particles.
In order to thermalize, these particles would have to gain energy
through multibody scatterings to a  smaller
number of higher energy particles.
However, baryon number conservation prevents this,
thereby limiting the actual reheat temperature to
$T_R \sim m_{\phi}$.
Once the decay products do thermalize, the plasma can carry
at most roughly one unit of baryon number per degree of freedom,
giving $n_b/s \sim {\cal O}(1)$ \cite{lindead}.
In this large entropy case,
the production of such a degenerate plasma can also be
couched in terms of a chemical potential for baryon number \cite{mu}.
For an acceptable baryon to entropy ratio to result
the $B$ violating parameters of the soft potential must be suppressed,
or there must be an additional source of entropy at or below the
electroweak phase transition.

As given in the appendix, the directions $QQQL$
and $\bar{u} \bar{u} \bar{d} \bar{e}$ carry $B-L=0$ but have nonzero
$B+L$.
If baryogenesis takes place along one of these directions with
$T_R$ greater than $T_c$, the  temperature of the electroweak phase
transition,
anomalous sphaleron processes destroy the generated $B+L$ \cite{krs}.
It has been suggested that the condensate can survive to a temperature
below $T_c$, thereby suppressing the sphalerons if $\phi \sim T_c$
at $T_c$, and allowing baryogenesis along directions which carry
$B-L=0$ \cite{mu}.
However, as discussed above, the condensate
decays at or above  this scale.
So it may be marginally possible for baryogenesis to take place
along $B-L=0$ directions in the $W=0$ scenario.
Baryogenesis can certainly take place along exact flat directions
with $B-L \neq 0$ \cite{ad}.
The main drawback here is the requirement for additional entropy
releases below the electroweak phase transition.

\section{Cosmological Evolution of String Moduli}

\label{stringevol}

String moduli are
exactly flat (perturbatively) in the supersymmetric limit, and couple
to standard model fields only through Planck scale suppressed
interactions.
The coherent production of string moduli leads to the string
version \cite{bkn,rt} of the Polonyi problem \cite{polonyi}.
Such a condensate decays at a very low temperature, $T \sim 5$ keV,
and leads to a number of cosmological problems.
These include modification of the light element
abundances \cite{polonyi},
the requirement for baryogenesis at such a low temperature,
and overproduction of LSPs \cite{rt}.
In discussions of the cosmological evolution of string moduli,
it is usually assumed that $V^{\prime \prime} \sim \mgravitino^2$
at very early times.
If this were the case, the moduli would effectively be frozen
for $H \gg \mgravitino$, and begin to oscillate when
$H \sim \mgravitino$.
If the initial displacement were ${\cal O}(M_p)$ the moduli
dominate the energy density essentially as soon as oscillations
begin, leading to the cosmological disasters mentioned above.

We have seen, however, that at early times
because of the finite density induced soft potential,
$V^{\prime \prime} \sim H^2$, so that the fields are
parameterically close to critically damped.
During inflation the fields are therefore driven to a local
minimum within a few $e$-foldings ( unless the  induced mass happens
to be numerically much less than $H$).
The induced potential of course remains after inflation.
However, in general the minima of the induced potential
do not necessarily coincide with the minima of the
low energy potential.
In fact since the scale for variations in the soft potential
is ${\cal O}(M_p)$, one expects the minima to differ by this amount.
Once $H \sim M_p$, the moduli start to oscillate freely about
a minimum of the low energy potential with initial amplitudes
of ${\cal O}(M_p)$. 

As an example of the moduli evolution consider the following toy model
\beq
V= (m_{3/2}^2 + a^2H^2) |\M|^2 + {1 \over 2 M_p^2}
(m_{3/2}^2 + b^2H^2) |\M|^4 .
\label{modulimodel}
\eeq
For $H \gg \mgravitino$, the minimum lies at $(a/b)M_p$.
For $H \ll \mgravitino$, the
minimum lies at $M_p$.
For suitable
$a$ and $b$, the system sits near the first minimum until
$H \sim \mgravitino$.
At this point, the field begins to oscillate about the second minimum,
with ``initial'' amplitude
of ${\cal M} \sim (1-a/b) M_p$.
This is, of course, just a statement of the original Polonyi
problem.
Our observation that the curvature of the potential is of order the
Hubble constant at early times,
rather than ameliorating the problem, just gives a concrete realization
of the initial conditions.

The present discussion suggests a solution of the moduli
problem \cite{adshort}.  
If the minima coincide at early and late times the moduli are
driven to a minimum during inflation (up to quantum
deSitter fluctuations).
This is technically natural if there is a point of enhanced symmetry
on moduli space.
Very roughly, the moduli transform under some symmetry near such points.
The lowest order invariants are therefore bilinears, and the potential
(no matter what the source)
is necessarily an extremum at such points.
So it is possible that the potential is a minimum at a symmetry point at
both early and late times.
More precisely the moduli are composite fields near points of enhanced
symmetry.
At the symmetry point, the fields making up the moduli become
massless (ignoring any nonperturbative effects).
In string theory, there often exist points of enhanced gauge
symmetry on moduli space.
The moduli act as Higgs fields near the symmetry points.
The most famous example of this is self dual point of
$R \rightarrow {2 \over R}$ duality in
toroidal compactification.
At such points the Kaluza Klein $U(1)$ for each $S^1$ gets
enlarged to $SU(2)$.
Analogous points seem to be a generic feature of many compactifications.

As an example of enhanced symmetry, consider first moduli other than
the dilaton.
For these, it is
possible in many instances to find points where all
the moduli transform under a discrete symmetry.
An example is provided by the $Z_3$ orbifold \cite{orbifold}.
This orbifold can be constructed as a product of three
two-dimensional tori, each exhibiting a $Z_3$ symmetry.
(One of these $Z_3$'s is modded out; the other two survive).
The enhanced symmetry at this point is $SU(3) \times Z_3
\times Z_3$.  All of the moduli in the twisted sectors
are charged under $SU(3)$.  Of the untwisted moduli,
all but $3$ transform under the $Z_3$'s; these correspond
to breathing modes for the three tori.
However, for particular values of the radii and of
the antisymmetric tensor fields (torsion), there are further
enhanced symmetries.  In particular, one can go to
what would be the $SU(3)$ points of conventional 2-d
toroidal compactifications, for each
of the three tori.  After modding out by
the $Z_3$, six $U(1)$'s remain, under which
all of the remaining moduli transform.

Clearly the $Z_3$ orbifold does not describe the real
world.  On the other hand, this example illustrates the
possibility that all of the moduli (except the dilaton)
can transform under some enhanced symmetry.
For this scenario to be realized in a realistic example
our vacuum must be at or near a point
of enhanced symmetry.
Any enhanced continuous gauge symmetries must either be
identified with part or all of the standard model gauge
group or be spontaneously broken.
In the latter case there would be extra light gauge bosons at
the weak scale.
Upon supersymmetry
breaking, it is perfectly possible that some of the
continuous gauge symmetries are broken by ${\cal O}(m_{3/2})$ vev's.

The main problem with this idea is the dilaton.
One might hope that $S$ duality could be realized in the
effective potential.
However, at the dilaton self dual point the four dimensional gauge
coupling is likely to be very large.
So if symmetries are the solution of the moduli problem,
the dilaton is probably on different footing than the
other moduli.
The dilaton mass might arise from dynamics not directly associated
with supersymmetry breaking \cite{bkn,rt,moduliinfb}.
In no-scale type theories it may also be possible in some
circumstances to avoid the dilaton problem if the dilaton
dominates the energy density at very early times
\cite{counnas}.
Alternately a period of late inflation can in principle dilute the
dilaton \cite{rt,lsgut}.

\section{Conclusions}

Exact and approximate flat directions are a generic feature
of supersymmetric theories.
If low energy supersymmetry
has anything to do with nature, these flat directions
are likely to play an important role in cosmology.
The coherent production of scalar fields along flat directions
emerges as a generic feature of supersymmetric theories.
In this paper, we have explored certain aspects of the cosmology of
flat directions.
Perhaps our most important, albeit quite simple, observation is
that the scale for the induced soft potential
at early times is of order $H$.
This has dramatic consequences for the
AD mechanism of baryogenesis, and the evolution of string
moduli.

The AD mechanism of baryogenesis is not generally obtained
with a minimal Kahler potential for the standard model fields.
In this case the induced potential has a minimum at the origin,
and the fields are driven to small values during inflation.
With nonminimal Kahler couplings the origin can be unstable and
large expectation values along flat directions can result.
Such nonminimal couplings can in fact be generated radiatively
in the presence of Yukawa couplings.
For directions lifted by nonrenormalizable terms in the superpotential,
the fields begin to oscillate freely
in the low energy potential
at $H \sim \mgravitino$ with an ``initial'' condition which
is determined by a balance between
the induced soft mass term with and
nonrenormalizable term.
This guarantees that the $B$ conserving and violating terms in the
potential are the same order.
The resulting baryon number per condensate particle is near maximal.
However, since the ``initial'' field value is parameterically less
than $M_p$, the baryon to entropy ratio can be quite small.
Within this scenario the mechanism is quite robust.
For the $LH_u$ direction an acceptable baryon number results
with a reasonable value for the reheat temperature after inflation.
In this case the baryon asymmetry is related to the lightest
neutrino mass.

For the moduli problem, the induced potential generally gives
a concrete realization of the initial conditions, which are
usually just assumed in an ad hoc way.
However, it suggests a solution if there is an enhanced symmetry
point on moduli space.
The minimum of the potential can then in principle coincide at
early and late times.
However, it is not clear how the dilaton can fit into such a
picture.
If symmetries are the solution of the moduli problem,
our vacuum is quite close to an enhanced symmetry point,
which might have interesting phenomenological consequences.

\section{Acknowledgments}
L. R. would like to thank Bob Singleton for sharing his Mathematica
expertise, and Greg Anderson, Diego Castano, Dan Freedman,
Steve Martin, and Graham Ross for useful conversations. We would
also like to thank Lawrence Hall,
Hitoshi Murayama and Joel Primack for useful comments.

\section{Appendix}

In the global limit the scalar potential is a sum of gauge and
superpotential contributions,
$V=\frac{1}{2}g^2 D^a D^a + F_{\varphi}^* F_{\varphi}$, where
$D^a = \varphi^* T^a \varphi$,
and $F_{\varphi}^* = \partial W / \partial \varphi$.
There are many flat directions in the standard model field space
on which the potential vanishes with respect to the renormalizable
superpotential. $D$ flat directions can be parameterized
through gauge invariant operators.
In order to form such invariants
it is useful to first construct
 potentially $D$, and $F$ flat combinations of fields.
A set of such  operators which are $F$ flat with respect to
the standard model Yukawa couplings
$$
W = \lambda_u Q H_u \bar{u}
   + \lambda_d Q H_d \bar{d}
   + \lambda_e L H_d \bar{e}
$$
are given in table \ref{table1}.
Throughout, unbarred fields are $SU(2)_L$ doublets while
barred fields are $SU(2)_L$ singlets, and
generation indices are suppressed.
The Higgs fields appear in only a limited number of flat
directions.
The field $H_d$ does not appear in table \ref{table1} contracted with
$Q$ or $L$ since the Yukawa couplings
give
nonzero $F_{\bd}^*$ and $F_{\be}^*$ respectively in this case.
Likewise $H_u$ does not appear contracted with $Q$ as
$F_{\bu}^*$ would be nonzero.
It can appear contracted with $H_d$ or $L$ though.
The only flat directions involving Higgs fields are
therefore
$H_u H_d$ and $H_u L$.
The superpotential $\mu$ term,
$$
W = \mu H_u H_d
$$
generates a nonzero
$F$ component if either Higgs is nonzero,
$F_{H_u}^* = \mu H_d$ and $F_{H_d}^* = \mu H_u$.
However $\mu$ can not be much larger than the weak scale.
So this contribution to the potential is the same order
as that from the zero density soft supersymmetry breaking masses.
Directions which are lifted only by the $\mu$ term in the
supersymmetric limit are therefore included in the list
of renormalizable ``flat'' directions.

Flat directions can be constructed by tensoring
together products of the
combinations of fields that appear in table 1.
A complete list of standard model flat directions
made out of up to 7 fields 
is given in table 2.
Gauge indices are contracted in an obvious manner.
In general many possible
directions exist for each invariant by permuting flavor indices.
For example, the invariant $LL \bar{e}$
contains the independent $F$ flat invariants
$L_1 L_2 \be_3$,
$L_1 L_3 \be_2$, and
$L_2 L_3 \be_1$.
For each invariant there is a single
Goldstone boson
for the spontaneously
broken $U(1)$ global quantum number carried by the invariant,
and its supersymmetric partner.
The scalar components of the other directions in the invariant are Goldstone
bosons for the spontaneously broken flavor symmetries.

Fields can take values along multiple directions
simultaneously, and remain $F$ flat, although this is not
guaranteed.
For example, the directions $Q_1 L_1 \bar{d}_2$,
$\bar{u}_2 \bar{d}_2 \bar{d}_3$, and $L_1 L_2 \bar{e}_3$
can be nonzero and preserve the $D$ and $F$ flat conditions.
Typically most directions not related by flavor
are lifted when fields take on nonzero value along some direction.
For example, when $LH_u$ is nonzero, all the directions
in table \ref{table2} are lifted except
$LL\be$ and $LL\bd \bd \bd$.

All the flat directions in table 2 made of given number
of fields have the same $B-L$, with the exception
of $H_u H_d$ and $LL \bd \bd \bd$.
This may seem surprising, but follows from the following
simple combinatorics.
 As listed in table 1 the number of fields minus $B-L$
equals 4 for units which do not involve
Higgs fields, and is zero for $\bar{e}$.
This has the effect that all directions
made out of a given number of
$F$ flat combinations from table 1,
any number of $\be$ fields, and no Higgs fields,
have the same $B-L$.
All the directions listed in table 2
without Higgs fields 
are in fact constructed in this way,
with the exception of $LL \bd \bd \bd$.
This direction is made of two operators from table 1 while
the other directions with 5 fields are made from
one operator and some number of $\be$ fields.

\newpage

\begin{table}[h]
$$
\begin{array}{lrrc} \hline \hline
 & & &  \\
 &  \multicolumn{1}{c}{Y}  &
    \multicolumn{1}{c}{B-L} &
    \multicolumn{1}{c}{N - (B-L)} \\
 & & & \\
\hline
 & & & \\
\bar{e}  & 2  &  1 &  0 \\
 & & & \\
L L       & -2 & -2 &  4 \\
 & & & \\
H_u H_d &  0  & 0  &  2 \\
 & & & \\
L H_u   &  0  & -1 &  3 \\
 & & & \\
\bar{u} \bar{u} \bar{u}
         & -4 & -1 &  4 \\
 & & & \\
\bar{u} \bar{u} \bar{d}
         & -2 & -1 &  4 \\
 & & & \\
\bar{u} \bar{d} \bar{d}
         & 0  & -1 &  4 \\
 & & & \\
\bar{d} \bar{d} \bar{d}
         & 2  & -1 &  4 \\
 & & & \\
Q L \bar{u}
         & -2  & -1 &  4 \\
 & & & \\
Q L \bar{d}
         & 0  & -1 &  4 \\
 & & & \\
Q Q \bar{u} \bar{u}
         & -2 &  0 &  4 \\
 & & & \\
Q Q \bar{u} \bar{d}
         &  0 &  0 &  4 \\
 & & & \\
Q Q \bar{d} \bar{d}
         &  2 &  0 &  4 \\
 & & & \\
Q Q Q L
         &  0 &  0 &  4 \\
 & & & \\
QQQQ\bu
         &  0 &  1 &  4 \\
 & & & \\
QQQQ\bd
         &  2 &  1 &  4 \\
 & & & \\
Q^6
         &  2 &  2 &  4 \\
& & & \\
\hline
\end{array}
$$
\caption{ Combinations of fields for
constructing flat directions.}
\label{table1}
\end{table}

\newpage

\begin{table}[h]
$$
\begin{array}{lr}\hline \hline
 & \\
 & \multicolumn{1}{c}{B-L}
     \\  \hline
 & \\
H_u H_d    &  0  \\
L H_u      & -1  \\
 & \\
\bar{u} \bd \bd          & -1  \\
Q L \bd                  & -1  \\
L L \be                  & -1  \\
 & \\
QQ \bu \bd               & 0  \\
QQQL                     & 0  \\
QL \bu \be               & 0  \\
\bu \bu \bd \be          & 0  \\
 & \\
QQQQ\bu                  & 1   \\
QQ \bu \bu \be           & 1  \\
LL\bd \bd \bd            & -3  \\
\bu \bu \bu \be \be      & 1   \\
 & \\
QLQL \bd \bd             & -2  \\
QQLL \bd \bd             & -2  \\
\bu \bu \bd \bd \bd \bd   & -2 \\
 & \\
QQQQ\bd LL               & -1  \\
QLQLQL \be               & -1   \\
QL \bu QQ\bd \bd         & -1  \\
\bu \bu \bu \bd \bd \bd \be & -1 \\
& \\
\hline
\end{array}
$$
\caption{Renormalizable $F$ and $D$ flat directions in the
standard model.}
\label{table2}
\end{table}

\newpage

\large
\noindent
{\bf Figure Captions}
\vskip.5in
\normalsize

\begin{description}

\item[Figure 1.] Trajectory in $\phi$ plane for
$n \theta_i = \frac{9}{10} \pi$, with $m_{\phi}=c^{\prime}=-a=1$, and
$n=4$.

\item[Figure 2.] Fractional baryon number carried by condensate
as a function of $n \theta_i$ for
$m_{\phi}=c^{\prime}=-a=1$, and $n=4$.

\end{description}


\begin{thebibliography}{20}
\bibitem{yukawanote}
The only exception is a Yukawa coupling of two fields in the flat
direction directly to a field in the inflaton sector with small
expectation value but large $F$ component.
The $F$ component would then induce
a soft mass for the flat direction of $m \sim g \sqrt{H M_p}$.
\bibitem{Dinf} Nonzero $D$ terms can arise during
inflation in the presence of a linear Fayet-Iliopoulos
$D$ term. For example, consider an $SU(2) \times U(1)$ model
with gauge couplings $g$ and $e$, and field content
$\varphi=(2,+1)$ and $\bar{\varphi}=(2,-1)$.
With a linear $D$ term for the $U(1)$, $\xi \intD V$,
there is flat direction
$\varphi^* \varphi - \bar{\varphi}^* \bar{\varphi} =
  - \frac{1}{2}(e^2/(e^2+g^2)) \xi$,
along which the gauge potential does not vanish,
$V_D=\frac{1}{8}( e^2 g^2 / (e^2 +g^2)) \xi^2$.
With a nonminimal Kahler potential, inflation can take place along
this direction.
\bibitem{phasenote} If no potential for the phase of $\phi$
exists during inflation the baryons which are eventually produced
have a (roughly) scale invariant spectrum of density fluctuations,
$\delta n_b / n_b \sim H_I / |\phi_0|$, where $|\phi_0|$
is the minimum during inflation.
These fluctuations can not be too large
in order that structure does not form too early after matter domination,
i.e. $\delta \rho_b / \rho_{\rm tot} < 10^{-5}$.
This gives a bound of $\phi_0 / H_I < 10^4$ during inflation.
\bibitem{LHnote}
If this operator is generated by integrating out
a heavy singlet with $m_N < H_I$, the singlet can be excited
during inflation with amplitude
$\langle \delta N^2 \rangle  \sim H_I^2$.
This gives an effective mass squared to the $LH_u$ direction of
$m^2 \sim g^2 H_I^2$. But to obtain an acceptable baryon number
in this case $g \ll 1$.
So the negative mass squared scenario can still be realized even if
the singlet is excited during inflation.
\bibitem{limitnote} Of course for very small $m_{\nu}$ the
baryon number must vanish as $m_{\nu} \rightarrow 0$.
For $m_{\nu} < 10^{-20}$ eV the $\frac{\lambda}{M}(LH_u)^2$ term
in the superpotential is less important in determining the
value of the $LH_u$ direction when $H \sim \mgravitino$
than higher order terms in the soft potential.
For $m_{\nu}$ less than this value the baryon number
therefore does decrease (ignoring any $U(1)_L$ violation in the
Kahler potential).
\end{thebibliography}
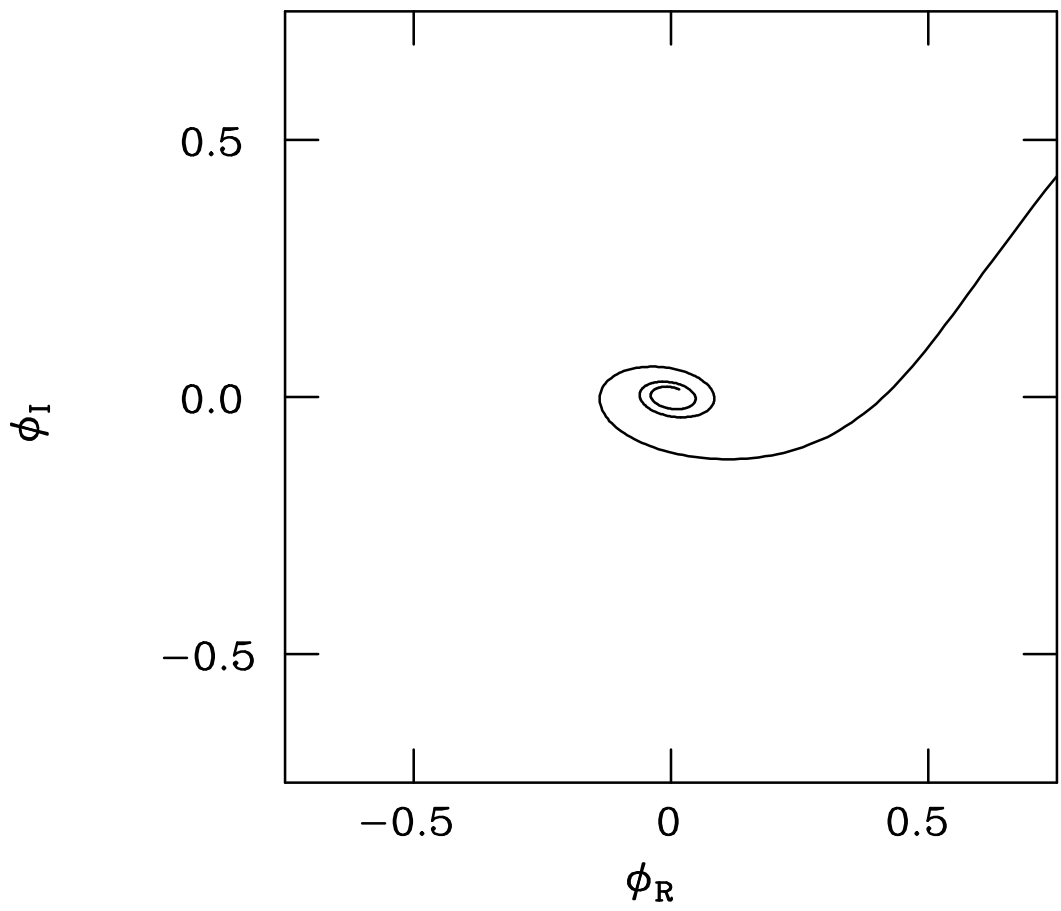
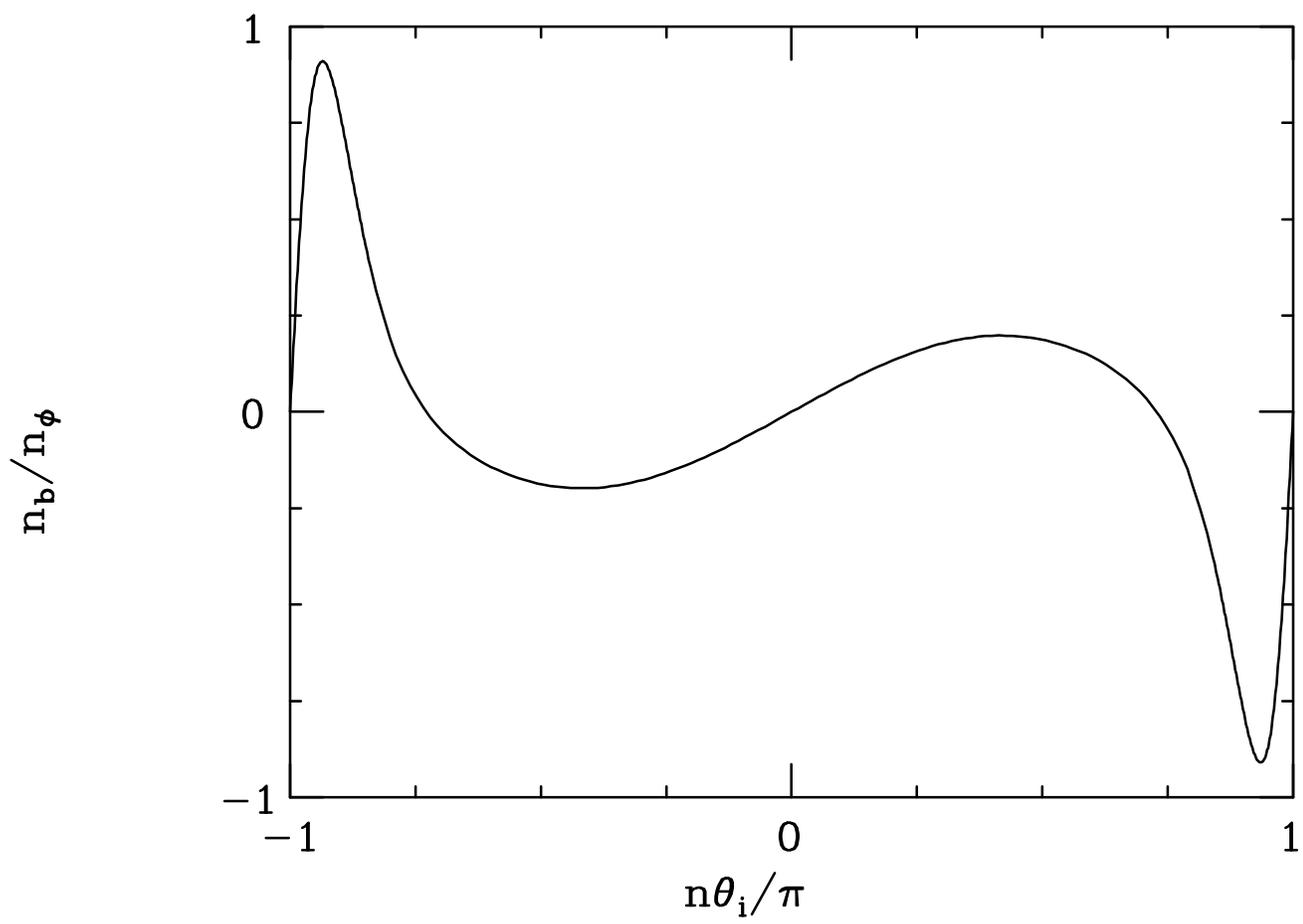
\end{document}